\documentclass[twocolumn, amssymb,  amsmath, amsfonts, nobibnotes, a4paper, aps, prd, superscriptaddress, nofootinbib]{revtex4-2}

\usepackage[T3,T1]{fontenc}
\DeclareMathAlphabet{\mathpzc}{OT1}{pzc}{m}{it} 
\usepackage{mathrsfs} 
\usepackage{bm} 
\makeatletter
\@ifclassloaded{beamer}
  { 
    \typeout{UsePackages: Detected beamer}
    \usepackage{tgheros}
    
  }
  { 
    \typeout{UsePackages: Did not detect beamer}
    \ifx\asybeamer\undefined 
    \typeout{UsePackages: Detected article}
    \usepackage[varg]{txfonts} 
    \usepackage{tgtermes} 
    \else 
    \typeout{Fonts: Detected asy for beamer}
    \usepackage{tgheros}
    
    \fi
  }

\usepackage{siunitx}
\usepackage{relsize}
\usepackage{booktabs}

\usepackage{multirow, newfloat}
\usepackage{graphicx} %
\usepackage[table,x11names,svgnames,rgb]{xcolor}
\usepackage{xspace}

\usepackage{mathtools}
\DeclareSymbolFontAlphabet{\mathrm}{operators}

\makeatletter %
\newcommand{\ShowFont}{%
  \typeout{The main font is \f@encoding \space \f@family \space %
    \f@series \space \f@shape \space at \f@size pt.}%
  \typeout{The math font sizes are \tf@size pt (main), \sf@size pt %
    (script), and \ssf@size pt (scriptscript).}%
  \typeout{The linewidth is \the\linewidth}} %
\makeatother %

\usepackage[colorlinks]{hyperref}
\usepackage{orcidlink}

\newcommand{\Mc}{\ensuremath{\mathcal{M}_c}\xspace}
\def \match     {\mathscr{M} }

\usepackage[first=0,last=9]{lcg}

\DeclareRobustCommand{\modelstyle}[1]{\texttt{#1}}

\newcommand{\SEOBvFive}{\mbox{\modelstyle{SEOBNRv5\_ROM}}\xspace}

\newcommand{\pycbc}{\mbox{\texttt{PyCBC}}\xspace}
\newcommand{\lalsuite}{\mbox{\texttt{LALSuite}}\xspace}

\newcommand{\gravelamps}{\mbox{\texttt{Gravelamps}}\xspace}
\newcommand{\numpy}{\mbox{\texttt{NumPy}}\xspace}
\newcommand{\scipy}{\mbox{\texttt{SciPy}}\xspace}
\newcommand{\matplotlib}{\mbox{\texttt{Matplotlib}}\xspace}

\DeclareMathAlphabet{\mathcalstd}{OMS}{cmsy}{m}{n}
\DeclareMathAlphabet{\mathpzc}{OT1}{pzc}{m}{it}

\newcommand{\UCLouvain}{University of Louvain, Centre for Cosmology, Particle Physics and Phenomenology---CP3, \\ Chemin du Cyclotron 2, 1348 Louvain-la-Neuve, Belgium}
\newcommand{\ROB}{Royal Observatory of Belgium, Avenue Circulaire, 3, 1180 Uccle, Belgium}

\definecolor{dodgerblue}{HTML}{1E90FF}
\hypersetup{citecolor=dodgerblue,  urlcolor=dodgerblue}

\definecolor{RED}{HTML}{F5054F}
\definecolor{LIGHT_ORANGE}{HTML}{FDAA48}
\definecolor{DARK_ORANGE}{HTML}{FF5B00}
\definecolor{LIGHT_BLUE}{HTML}{448EE4}
\definecolor{DARK_BLUE}{HTML}{0343df}
\definecolor{LIGHT_GREEN}{HTML}{40C53C}
\definecolor{DARK_GREEN}{HTML}{02590F}
\definecolor{rowshade}{gray}{0.91}

\newcommand{\Msun}{\mathrm{M}_\odot}
\newcommand{\ML}{M_{\rm L}}
\newcommand{\murel}{\mu_{\rm rel}}
\newcommand{\dt}{\Delta t}
\newcommand{\chisq}{\chi^2_{\rm r}}
\newcommand{\chisqL}{\chi^2_{{\rm r, L}}}
\newcommand{\newsnr}{\hat{\rho}}
\newcommand{\rhoL}{\rho_{\rm L}}
\newcommand{\Ffac}{F(f)}
\newcommand{\fml}{f_{\rm ML}}

\begin{document}

\title{Efficient search for microlensed gravitational-wave signals with matched filtering and a two-image statistic}

\author{Soumen Roy\orcidlink{0000-0003-2147-5411}}
\email{soumen.roy@uclouvain.be}
\affiliation{\UCLouvain}
\affiliation{\ROB}

\author{Justin Janquart\orcidlink{0000-0003-2888-7152}}
\affiliation{\UCLouvain}
\affiliation{\ROB}

\date{\today}

\begin{abstract}
Gravitational waves from compact binary coalescences can be microlensed by compact objects along the line of sight, producing frequency-dependent waveform distortions. As standard matched-filter searches rely on unlensed templates without accounting for lensing effects, microlensed signals can lose signal-to-noise ratio (SNR) through template mismatch and suffer further losses from signal-consistency tests. We quantify both effects for the \texttt{PyCBC} search using microlensed injections in simulated data. The two losses occupy different regions of lens parameter space in the geometric optics representation: template mismatch is governed by the relative magnification of the two images, while the consistency-test penalty is governed by their time delay. We show that a microlensed signal lies in the two-dimensional subspace spanned by a template and its time-shifted copy, allowing the relative magnification and Morse phase to be maximized over analytically, leaving only the time delay to search. The resulting statistic requires no extension of the template bank and reduces the median loss in re-weighted SNR from $10\%$ to $1.6\%$, and the $90^\mathrm{th}$ percentile from $41\%$ to $9\%$. At a false-alarm rate of one per day, the relative detection volume improves by about $6\%$ across all injections, and by a factor of $1.9$ for the $10\%$ of injections most affected by lensing. In addition to increased detection capacities, we also show how the detection statistic can be used to establish the lensed nature of the trigger. Therefore, the work proposed here lays the foundation for an effective search and identification of microlensed signals in gravitational-wave data.
\end{abstract}

\maketitle

\section{Introduction}
\label{sec:intro}

Gravitational waves (GWs) propagating through the Universe can be deflected by intervening mass distributions. The effect depends on the relative size of the lens compared with the GW wavelength. A galaxy or a cluster is far larger than the wavelength, and it produces repeated images of the same source that arrive as separate events, minutes to months apart. A lens of redshifted mass $\ML$ in the range $10$--$10^5\,\Msun$ is comparable to the wavelength, and it instead produces a single distorted signal. Depending on the impact parameter $y$, the distortion is a frequency-dependent beating pattern in the wave-optics regime, or a sum of overlapping repeated images in geometric optics~\cite{Takahashi:2003ix, Nakamura:1997sw}. So far, no confident signature of a lensed signal has been found in LIGO--Virgo--KAGRA (LVK) data~\cite{Hannuksela:2019kle, LIGOScientific:2021izm, LIGOScientific:2023bwz, Janquart:2023mvf, LIGOScientific:2025cwb}.

In this work, we focus on the case where lensing leads to a single distorted signal. Therefore, the detection problem corresponds to the identification of a trigger and the assessment of its lens nature. Conversely, for strong lensing searches, one would need to search for multiple images corresponding to the same signal~\cite{Liu:2020par, Lo:2021nae, Janquart:2021qov, Janquart:2023osz}. Some of the methods developed here can be extended to that case or other search methods can be used to verify signal coherence~\cite{Heynen:2025fgq}. Here, we do not address the multiple-signal issue. 
The lensing regime probed here is unique to GWs since, for electromagnetic signals, such effects correspond to  a simple change in brightness and are hard to measure. For GWs, the information on the lens itself is encoded in the frequency-dependent distortions imprinted on the detected GW signal. Searching for this signature is therefore of interest because it probes the abundance of compact objects in a mass range that is difficult to access otherwise and gives information about the nature of dark matter~\cite{Lai:2018rto, Diego:2019rzc, Urrutia:2021qak, Tambalo:2022wlm}. 

Searches for compact binary coalescences (CBCs) in LVK data are mostly based on matched filtering against banks of template waveforms computed in general relativity for unlensed, quasicircular binaries~\cite{Usman2016, Allen2012, Messick2017, Aubin2021}, and searches for lensed signals usually start from the triggers that such searches produce~\cite{Wright:2025nod}. A microlensed signal is therefore filtered with a template that does not describe it, and two distinct consequences follow. The first is familiar: no template in the bank matches the distorted waveform, so the recovered matched-filter SNR is lower than the optimal value. The second is less widely appreciated. Searches do not rank candidates on SNR alone; they apply a signal-consistency test, which compares the distribution of matched-filter power across the band with that expected from the recovered template, and down-rank candidates that disagree~\cite{Allen2005, Messick2017}. Such a test has as main goal to discard instrumental noise. However,  since a distorted lensed signal has its time-frequency evolution modified, there is a disagreement with the template and the trigger may be discarded, even though its origin is astrophysical.

The two effects (SNR and consistency test) are not equally recoverable. A signal that loses SNR to mismatch is generally ranked as a signal, and a follow-up analysis can recover it. A signal that fails the consistency test is ranked as a glitch, and it is demoted below the threshold at which candidates are kept. Because the lensing analyses start from the trigger lists of the unlensed searches, such a candidate is never examined. The signals affected are those whose two images, when approximating the microlensing regime in the geometric optics, overlap inside the detector band~\cite{Lo:2024wqm}.

The consequences of the signal consistency-test have been studied recently. Ref.~\cite{Chan:2024qmb} carried out an injection campaign with the \textsc{GstLAL} pipeline and found that a broad region of the lens parameter space is down-ranked by this unlensed search. Ref.~\cite{ChanLi2026} characterised the response of representative time- and frequency-domain consistency statistics analytically over the two-image parameter space $(\murel, \dt)$, identified the regime in which the penalty is reduced, and proposed an asymmetry statistic built from the time-domain residual to recover down-ranked candidates. Separately, extending the template bank to include lensed waveforms has been proposed~\cite{Gholap:2026fjy}. While dedicated banks would be the most optimal, such solutions are computationally the most expensive due to the increase in dimensionality.

What has not been done is an end-to-end study for a frequency-binned $\chi^2$ test with a realistic template bank. The distinction matters for two reasons. First, an analytic treatment that filters the signal with a template carrying the true source parameters cannot separate the SNR lost because the bank does not contain a matching template from the SNR removed by the consistency test, since the former is zero by construction. Second, a real search maximizes over a bank, and the template it selects for a distorted signal is displaced from the true parameters in a lensing-dependent way, which affects both quantities.

In this paper we measure both effects for the \pycbc\ search~\cite{DalCanton:2014hxh, Usman:2015kfa, Allen:2005fk, Nitz:2017svb}, using $10,000$ microlensed injections analyzed with an aligned-spin template bank constructed for LVK searches of O4 data~\cite{LIGOScientific:2025zdk}. We find that the two losses populate the lens parameter space differently and must be distinguished to characterize the search response. We then show that the lensed waveform lies in a two-dimensional linear subspace spanned by the recovered template and a time-shifted copy of itself, so that two of the three lensing degrees of freedom can be maximized over analytically. We then build a lensing detection statistic that does not require an extension of the template bank, and removes most of the consistency penalty. Finally, we show how the comparison between unlensed and lensed search outputs can be used to establish the lensed nature of an identified trigger, demonstrating an end-to-end approach for searching for single distorted lensed signals.

Sec.~\ref{sec:model} details the assumptions made on the lensing model and the impact it has on the detected signals. Sec.~\ref{sec:selection} presents the measured response of the unlensed search across the lens parameter space and separates the two loss mechanisms by performing a large injection campaign. Sec.~\ref{sec:statistic} constructs the dedicated two-image lensed statistic and Sec.~\ref{sec:recovery} reports its improved recovery of lensed signals. Sec.~\ref{sec:background} measures the background with time slides, and reports the sensitivity of the combined search. Sec.~\ref{sec:identification} shows how the search outputs can be used to confirm the lensed nature of a trigger and the expected performance of the proposed approach. We discuss the implications and next steps in Sec.~\ref{sec:conclusion}.

While this work was being completed, an independent study by Wadekar \textit{et al.}~\cite{Wadekar:2026xse} appeared, which constructs the matched-filter output of a two-image template from the SNR time series that the unlensed search already computes, similar to our two-image statistic. The two papers test the idea in different settings. They work with a single detector and a single non-spinning binary, placed on a grid of lens parameters, and they obtain the recovery and the consistency penalty from the expected statistic in Gaussian noise. The template is then optimized continuously over the intrinsic parameters at the best-fit lens point, which, as they note, favours the unlensed search over the two-image one. We instead inject a population of binaries into simulated data and recover them with a full search in a two-detector network, using one shared time delay across the network and a pooled network statistic, and we re-maximize jointly over the template bank and the image time delay, making our setting closer to how a real search would be carried out. Our numbers therefore carry the discreteness of the bank and the clustering and thresholds of the search, and we measure the background with time slides on the same data, so that the quoted sensitivity gains include the trials factor of the delay search. We also show how search outputs can be used to test the lensing hypothesis, going beyond the identification of triggers. Where the two analyses overlap, the conclusions agree.

\section{Microlensed waveform}
\label{sec:model}

In the thin-lens, weak-field approximation the lensed waveform is related to the
unlensed one by a frequency-dependent amplification
factor ($F(f; \phi)$)~\cite{Takahashi:2003ix},
\begin{equation}
  \tilde{h}_L(f) = F(f; \phi)\, \tilde{h}(f) ,
  \label{eq:amp}
\end{equation}
where $\phi$ represent the lensing parameters. The amplification factor is obtained from the diffraction integral over the lens plane and depends on the lens model under consideration.

For the case of an isolated point-mass lens, as considered in this work, the lensing parameters correspond to the redshifted lens mass $M_L$ and the dimensionless impact parameter $y$, corresponding to the source position in units of Einstein radii.
It is then often convenient to work with the dimensionless frequency
\begin{equation} \label{eq:w}
  w = \frac{8 \pi G M_L f}{c^3} \, ,
\end{equation}
which compares the typical size of the lens with the wavelength of the gravitational wave. Depending on $w$'s value, we can distinguish different regimes. When $w \leq 1$, the wavelength of the gravitational wave is larger than the lens size, and we are in the wave-optics regime. In this case, diffraction effects are significant, and the amplification factor exhibits oscillatory behavior as a function of frequency. Conversely, when $w \gg 1$, the wavelength is much smaller than the lens size, and we enter the geometric optics regime, where multiple images with the same time-frequency evolution are formed. 

\subsection{Wave-optics regime}
In the wave-optics regime, the amplification factor for a point-mass lens can be expressed in terms of the confluent hypergeometric function $_1F_1$~\cite{Takahashi:2003ix}:
\begin{equation}
  \begin{split}
    F(w; y) = {}& \exp\left[\frac{\pi w}{4}
      - \frac{i w}{2}\left(\ln\frac{w}{2} - 2\phi_m(y)\right)\right] \\
    &\times \Gamma\!\left(1 + \frac{i w}{2}\right)
      \,{}_1F_1\!\left(-\frac{i w}{2},\, 1;\, -\frac{i w y^2}{2}\right) ,
  \end{split}
  \label{eq:amp_wave_optics}
\end{equation}
where $\Gamma$ is the gamma function and $\phi_m(y) = (x_m-y)^2/2 - \ln x_m$, with $x_m = (y+\sqrt{y^2+4})/2$, is the Fermat potential at the brighter image. The amplification factor in this regime exhibits oscillatory behavior as a function of the frequency. The additional phasing and amplitude due to lensing depend on ($y$, $M_L$) and can lead to significant modifications of the observed waveform, especially for good lens-source alignment.

Since the condition $w \propto f$ and the signal is chirping, it can happen that the condition $w \sim 1$ is met at some point during the signal. In that case, only a part of the signal is in the wave-optics regime, while the rest can be described using the geometric optics approximation. 

\subsection{Geometric-optics regime}
In this regime, when $w \gg 1$, the diffraction integral for the amplification factor is dominated by the stationary points of the time delay function. This leads to the description of the lensing amplification factor in terms of a sum of images of the original signal, each undergoing a time delay, a magnification, and an overall phase shift.~\cite{Takahashi:2003ix}:
\begin{equation}
    F(f; y) = \sum_{j} |\mu_j|^{1/2} \exp\left[- 2\pi i f t_j + i \pi n_j\right] \, ,
    \label{eq:amp_geo_optics}
\end{equation}
where the sum is over the images $j$ produced by the lens, $\mu_j$ is the magnification of the $j$-th image, $t_j$ is the time delay of the $j$-th image, and $n_j$ is the Morse index associated with the $j$-th image. The Morse index admits three discrete values: $\{0, 0.5, 1\}$, matching a minimum, saddle point, or maximum of the time delay function, respectively.

In the case of a point-mass lens, there are two images formed, corresponding to a minimum and a saddle point of the time delay function. Therefore, the amplification factor, using the appropriate convention for the Fourier transform~\cite{WrightFourierConvention}, can be expressed as
\begin{equation}
  \Ffac = \sqrt{|\mu_+|} + i \sqrt{|\mu_-|}\, e^{-2\pi i f \dt} \, ,
  \label{eq:Fgeo}
\end{equation}
where the magnifications ($\mu_+$ and $\mu_-$) and time delay ($\dt$) are functions of the impact parameter and the redshifted lens mass:
\begin{equation}
  \mu_\pm = \frac{1}{2} \pm \frac{y^2+2}{2y\sqrt{y^2+4}} \, ,
  \label{eq:mags}
\end{equation}
and 
\begin{equation}
  \dt = \frac{4 G \ML}{c^3}
        \left[ \frac{y\sqrt{y^2+4}}{2}
             + \ln \frac{\sqrt{y^2+4}+y}{\sqrt{y^2+4}-y} \right] \,. 
  \label{eq:dt}
\end{equation}

Normalizing the amplification factor by the brighter image's magnification leads to the more commonly used form~\cite{ChanLi2026, Blandford1986, Ezquiaga:2025gkd} 
\begin{equation}
  \Ffac = 1 + \murel \, e^{-i(2\pi f \dt - \pi/2)} ,
  \qquad
  \murel = \sqrt{|\mu_-|/|\mu_+|} ,
  \label{eq:Frel}
\end{equation}

In this work, we use the two-image representation of the lensing effect to derive a specific statistic to detect microlensed GW signals. Therefore, it is important to note certain properties of the amplification factor in Eq.~\eqref{eq:Frel}, which relate to our observations. First, the time delay depends on $\ML$ and $y$ together: for small $y$ the bracket in Eq.~\eqref{eq:dt} tends to $2y$, so contours of constant $\dt$ are lines of slope $-1$ in the $(\log \ML, \log y)$ plane, and a heavy lens with poor alignment produces the same delay as a light lens with tight alignment. The relative magnification (Eq.~\eqref{eq:Frel}), by contrast, depends on $y$ alone, so contours of constant $\murel$ are horizontal. Neither quantity is a function of the lens mass by itself, and we will see that the response of the search is organized by both. Additionally, this two-image form sets a natural frequency scale for the signal. The interference fringes in $|\Ffac|$ are separated by
\begin{equation}
  \fml = 1/\dt ,
  \label{eq:fml}
\end{equation}
and the search can respond only when $\fml$ falls inside the sensitive band and is resolved by whatever structure the statistic imposes on that band.

As mentioned above, the region of validity of the geometric-optics approximation depends on a comparison between the mass of the lens and the frequency of the signal. Therefore, representing the lensing effect in terms of two images may not always be valid. For example, for $\ML = 400$ $M_\odot$, the geometric-optics approximation is valid for frequencies above $\sim 20$ Hz, while for $\ML = 8 M_\odot$, geometric optics are valid for frequencies above $\sim 1000$ Hz. Thus, for certain regimes, the model used in the search may not be exact, but, as we will see, even when the signal is affected by wave-optics effects, a two-image model is able to capture the main features of the lensing effect. 

Throughout this work, injected microlensed signals are generated in the wave-optics regime. Therefore, we use \gravelamps~\cite{Wright:2021cbn} to compute the amplification factor in the wave-optics regime, and switch to the geometric optics description from Eq.~\eqref{eq:Fgeo} above $w=1000$, since the two regimes agree well in this limit.


\section{Sensitivity of an unlensed template bank search to microlensed signals}
\label{sec:selection}

\subsection{Search configuration}
\label{sec:config}
We consider the parameter space of microlensed binary black hole (BBH) signals with component masses uniform in $m_i \in [10,50]\,\Msun$, dimensionless aligned spins uniform in $\chi_i \in [-0.99, 0.99]$, isotropic sky positions and orientations, and lensing parameters log-uniform in \mbox{$\ML \in [10,10^5]\,\Msun$} and $y \in [0.01,2]$. These ranges are chosen to sample the response of the search across the lens parameter space. They do not represent an astrophysical population of lenses. The gains we report are therefore a function of the lens parameters, and not a population average. We set the luminosity distances so that the optimal network SNR is uniform in $[8, 50]$. This ensures that the set samples cover the full range of detectability rather than concentrating near threshold. Since the $\chi^2$ statistic grows as the square of the SNR for a fixed waveform mismatch, an injection campaign at fixed SNR would not reveal how the penalty varies across the detectable population.
\begin{table}[t]
\centering
\begin{ruledtabular}
\begin{tabular}{lcc}
Parameter & Range & Distribution \\
\hline
Component masses ($m_{1,2}$) & $[10,50]\,\Msun$ & Uniform \\
Aligned spins ($\chi_{1,2}$) & $[-0.99,0.99]$ & Uniform \\
Sky position & -- & Isotropic \\
Binary orientation & -- & Isotropic \\
Lens mass ($\ML$) & $[10,10^5]\,\Msun$ & Log-uniform \\
Impact parameter ($y$) & $[0.01,2]$ & Log-uniform \\
\end{tabular}
\end{ruledtabular}
\caption{Parameter ranges and prior distributions used for the simulated microlensed signals.}
\label{tab:ParamSpace}
\end{table}

The base waveform is \SEOBvFive~\cite{Pompili:2023tna}, multiplied by $\Ffac$ and then transformed to the time domain. A frequency-domain waveform is zero below $f_{\rm low}$ and discontinuous at the first bin, and the inverse Fourier transform of that edge is a broadband transient at the moment the signal switches on. We therefore condition the waveform following the same approach as \texttt{XLALSimInspiralTD} in \lalsuite~\cite{lalsuite}, which \pycbc\ also uses. We size the array from a spin-aware duration estimate, and we place the merger one inspiral duration from the start rather than at the centre. The array then holds the inspiral before the merger and the delayed image after it, and neither wraps around.
\begin{figure*}[t]
  \includegraphics[width=0.95\textwidth]{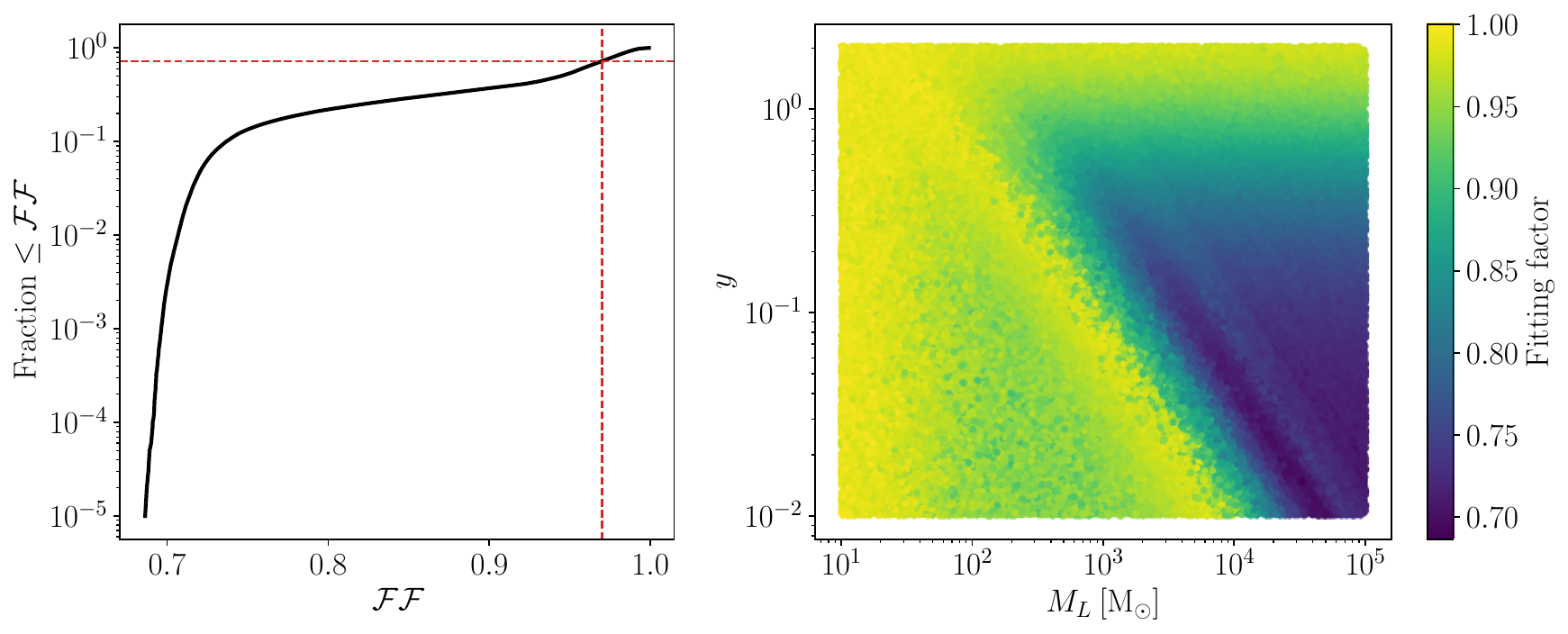}  
  \caption{Fitting factors obtained by filtering $10^5$ microlensed signals against the O4 unlensed template bank. The left panel shows the cumulative distribution of fitting factors, with the red dashed line indicating $\mathcal{FF}=0.97$. The right panel shows the fitting factors over the lens parameter space $(M_L, y)$, where each point represents an injection. The onset of the fitting factor loss follows a diagonal structure determined by the image time delay, while the magnitude of the loss depends on the relative magnification of the two images.}
  \label{fig:ff}
\end{figure*}

\subsection{Fitting factor}
\label{sec:ff}
To quantify the effectualness of the O4 unlensed template bank~\cite{LIGOScientific:2025zdk} in detecting microlensed signals, we carry out Monte Carlo simulations to compute fitting factors for a large set of injections. The template bank was constructed using a hybrid geometric-random placement algorithm~\cite{Roy:2017qgg, Roy:2017oul}, assuming the projected LIGO O4 sensitivity~\cite{LVK:2022NoiseCurves}. The fitting factor quantifies the fraction of optimal SNR that can be recovered by the template bank. Its distribution also helps identify regions of parameter space where the bank provides poor coverage.

For this study, we generate $10^5$ microlensed signals using the \SEOBvFive waveform model, with intrinsic parameters drawn uniformly from their respective ranges listed in Table~\ref{tab:ParamSpace}. The lens parameters are drawn from log-uniform distributions, as discussed in Sec.~\ref{sec:config}. For the non-precessing, dominant-quadrupole waveforms considered here, the fitting factors are independent of extrinsic parameters such as sky location, polarization angle, and inclination. This property also holds for the corresponding microlensed signals.

The fitting factor ($\mathpzc{FF}$) of a template bank ($\mathcal{T}$) with respect to an arbitrary signal ($h_a$) is defined as the maximum match over all the templates in the bank~\cite{Apostolatos:1995pj},
\begin{equation}
    \mathpzc{FF}(h_a) = \max_{\vec{\lambda} \in \mathcal{T}} \match(h_a, h(\vec\lambda)),
    \label{fittingfactor}
\end{equation}
where $\vec{\lambda}$ denotes a point in the template bank and the match is maximized over the relative time and phase. The fractional loss in the optimal SNR due to the mismatch between the signal and the template bank is therefore given by ($1-\mathcal{FF}$).

For each injection, we restrict the templates to a chirp-mass window around the injected chirp mass, $\mathcal{M}_c^{\rm inj}$. We consider a window of $\pm 30\%$ for $\mathcal{M}_c^{\rm inj} \leq 20\,\Msun$ and $\pm 50\%$ for $\mathcal{M}_c^{\rm inj} > 20\,\Msun$. These windows are chosen to cover the range of chirp masses of the best-matching templates, including the shifts introduced by microlensing. We further use the same selection criteria to construct a reduced O4 template bank containing 80,053 templates, which is used for the search analysis presented later in this work.

Fig.~\ref{fig:ff} shows the fitting-factor results obtained using the O4 unlensed template bank. The left panel shows the cumulative distribution of fitting factors for all injections, while the right panel shows the fitting factors over the lens parameter space ($M_L, y$). Similarly to Ref.~\cite{Chan:2024qmb}, we find that the fitting factors decrease significantly in a region of the lens parameter space, indicating that the unlensed template bank cannot fully recover the SNR of these microlensed signals. The onset of this loss follows a diagonal structure in the ($M_L, y$) plane, rather than being determined by the lens mass alone.

This behavior can be understood from the time delay between the two lensed images, which depends on both the lens mass and impact parameter, as shown in Eq.~\eqref{eq:dt}. For small impact parameters, contours of constant time delay follow diagonal lines in the logarithmic lens parameter space. As the time delay increases, the two images become increasingly distinguishable by the matched filter, resulting in a reduction of the fitting factor due to a loss in SNR. For sufficiently large time delays, the overlap between the two images becomes negligible. In this limit, a single unlensed template can recover the brighter image but cannot simultaneously recover the SNR contributed by the second image. Assuming an exact match to the brighter image, the fitting factor approaches
\begin{equation}
  \mathpzc{FF} \longrightarrow \frac{1}{\sqrt{1+\murel^2}} \, .
  \label{eq:ffloor}
\end{equation}
Since the relative magnification, $\mu_{\rm rel}$, depends only on the impact parameter, the limiting fitting factor is independent of the lens mass. This expression gives $\mathcal{FF}\simeq0.711$ for $y=0.01$ and $\mathcal{FF}\simeq0.986$ for $y=2$.

The right panel of Fig.~\ref{fig:ff} shows that the fitting factors approach limiting values at large lens masses. The reduction is more significant for smaller impact parameters, where the two images have comparable magnifications. For larger impact parameters, the second image becomes relatively faint, and most of the SNR can still be recovered by the unlensed template bank. Thus, the image time delay determines the region where the fitting-factor loss becomes significant, while the relative magnification determines the magnitude of this loss.

\subsection{Data and search}
\label{sec:data}
To quantify the response of an unlensed search to microlensed signals, we add the injections to simulated Gaussian noise colored to the O4 design sensitivity~\cite{LVK:2022NoiseCurves} for the two LIGO detectors. We analyze them with \texttt{pycbc\_inspiral}~\cite{Usman:2015kfa, Allen:2005fk}. The search uses a low-frequency cutoff of $20$~Hz and an SNR threshold of $5.5$, and filters against the reduced O4 bank of $80\,053$ templates described in Sec.~\ref{sec:ff}.

Candidates are tested with the power $\chi^2$ of Ref.~\cite{Allen2005}, which partitions the template into $p$ frequency bins chosen such that each contributes equally to the expected SNR$^2$, and compares the observed per-bin contributions with that expectation. The number of bins is not fixed. The \pycbc\ configuration sets it from the peak frequency of the template~\cite{pycbc-config},
\begin{equation}
  p = 0.72\, f_{\rm peak}^{\,0.7} ,
  \label{eq:nbins}
\end{equation}
where $f_{\rm peak}$ is the peak frequency of the template. We note that $p$ is therefore template-dependent, ranging from $20$ for the heaviest templates in our range to $73$ for the lightest and most strongly spinning ones. Additionally, since the bins carry equal expected SNR$^2$ and a chirp accumulates most of its SNR at low frequency, the bins are not uniform in width: for a $(30, 30)\,\Msun$ template the lowest bins are $3$--$4$~Hz wide, while the highest spans several hundred hertz. Both of these properties are relevant for interpreting the results presented below.

Writing $\chisq = \chi^2/(2p-2)$ for the reduced statistic, triggers have their SNR
re-weighted in the standard way~\cite{Babak:2012zx},
\begin{equation}
\label{eq:newsnr}
\hat{\rho} = \begin{cases} \mathrm{SNR}, & \chi_r^2 \le 1,\\[6pt]
  \mathrm{SNR} \left[ \left(1+(\chi_r^2)^3\right)/2 \right]^{-1/6}, & \chi_r^2 > 1,
\end{cases}
\end{equation}
and the search ranks candidates on $\newsnr$.

We associate triggers with injections within $0.5$~s of the expected arrival time in each detector, computed from the geocentric coalescence time and the detector time delay for the injected sky position. When multiple triggers fall within this window, we retain the one with the highest SNR, along with its corresponding $\chi^2$, bin count, and template.

Network quantities are formed by adding the single-detector values in quadrature, and the network $\chi^2$ by pooling, \mbox{$\chi^2_{\rm net} = \sum_i \chi^2_i / \sum_i (2p_i - 2)$}, which is the statistic that would be obtained by filtering both data streams together~\cite{Usman2016}. The bin count $p$ varies between detectors through Eq.~\eqref{eq:nbins}, since the search can recover a different template in each, and averaging the reduced values instead would therefore weight the two detectors incorrectly.

The standard \pycbc\ search requires exact-template coincidence: triggers from the two detectors must correspond to the same template and have arrival times consistent with the inter-detector light-travel time. Imposing these conditions on our injection-associated triggers excludes $6\%$ of the injections, almost entirely due to the shared-template requirement, and gives a median SNR loss of $10.8\%$ among the remaining injections. Selecting triggers based on the re-weighted SNR rather than the SNR would instead underestimate the loss, as it preferentially selects triggers that are less penalized by the consistency test. We therefore select triggers based on the SNR, without imposing exact-template coincidence. Since the two detectors may recover different templates, we choose the template from the detector with the higher SNR and use it for both detectors in the two-image fits described in Secs.~\ref{sec:statistic} and~\ref{sec:recovery}.

\begin{figure*}[t]
  \includegraphics[width=\linewidth]{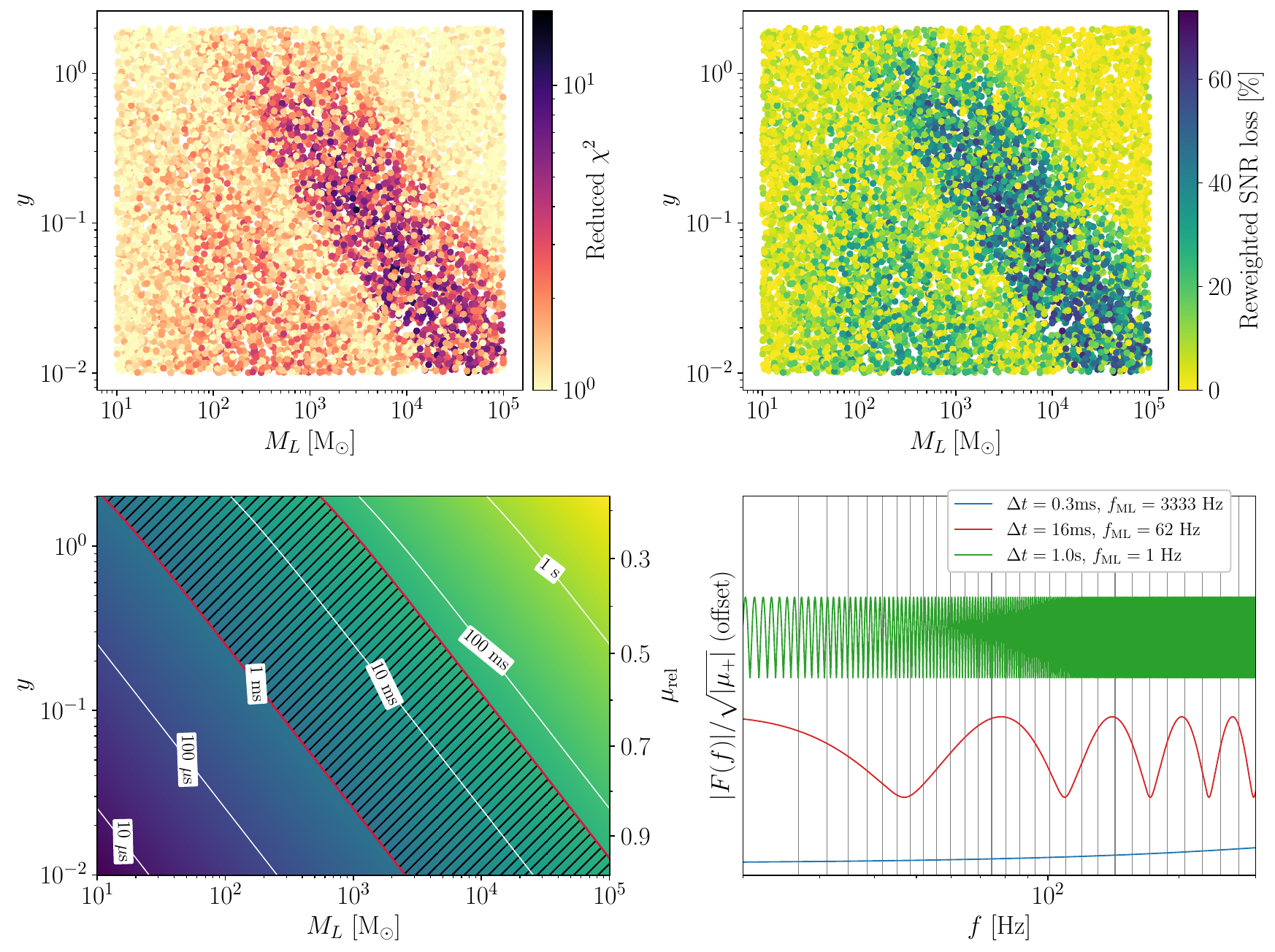}
  \caption{Response of the search to microlensed signals recovered using the reduced O4 unlensed template bank. The top panels show the measured network reduced $\chi^2$ (left) and the corresponding fractional loss in the re-weighted SNR, $(1-\newsnr/\rho)\times100\%$ (right). The bottom-left panel shows the image time delay over the same plane, with the relative magnification $\murel$ on the right-hand axis and the hatched strip marking $\fml = 1/\dt$ between $20$ and $1000$~Hz. The bottom-right panel shows the lensing amplification $|\Ffac|$ at $y=0.3$ for three delays, offset vertically, with the $\chi^2$ bin edges of a non-spinning $(30, 30)\:\Msun$ template drawn as vertical lines ($p=29$, $f_{\rm peak}=193.8$~Hz). The bins carry equal expected SNR$^2$ and therefore their widths range from $3.3$~Hz near $40$~Hz to $18$~Hz near $200$~Hz, which is what makes the response depend on where in the band the fringes fall.}
  \label{fig:lens_plane_loss}
\end{figure*}

\subsection{Loss from the consistency test}
\label{sec:chi2loss}

To interpret how a microlensed signal is affected by a standard search, we distinguish three losses. Writing $\rho_{\rm opt}$ for the optimal network SNR of the injected signal, $\rho$ for the network SNR recovered by the bank, and $\newsnr$ for the re-weighted value on which the search ranks candidates, we distinguish
\begin{align}
  1 - \rho/\rho_{\rm opt}
    &\quad \text{SNR loss due to the bank mismatch,} \label{eq:lossff}\\
  1 - \newsnr/\rho
    &\quad \text{SNR removed by the consistency test,} \label{eq:lossveto}\\
  1 - \newsnr/\rho_{\rm opt}
    &\quad \text{the two combined.} \label{eq:losstot}
\end{align}
Eq.~\eqref{eq:lossff} is the fitting-factor loss discussed in Sec.~\ref{sec:ff} and is a property of the bank, whereas Eq.~\eqref{eq:lossveto} is a property of the statistic. An analysis that filters the signal with a template carrying the true source parameters sets Eq.~\eqref{eq:lossff} to zero by construction and measures only Eq.~\eqref{eq:lossveto}. We measure both, and find that they behave differently over the lens parameter space.

Fig.~\ref{fig:lens_plane_loss} shows the network-reduced $\chi^2$ and the corresponding fractional loss in $\newsnr$. We find that the loss is concentrated in a diagonal band of the lens parameter space, reaching $73\%$ through it. We define the strip as the region with $1\,{\rm ms} < \dt < 50\,{\rm ms}$, equivalently $\fml$ between $20$ and $1000$~Hz, which is hatched in the bottom-left panel. It holds $44\%$ of the injections but $85\%$ of those losing more than $40\%$ of their SNR. This is the same band identified in the fitting factor, and it is diagonal for the same reason, since the image time delay depends on lens mass and impact parameter together. We find a median loss of $9.2\%$ over the full injection set, which rises to $16\%$ inside the strip. For comparison, the unlensed control set described in Sec.~\ref{sec:recovery} has a median loss of $0.6\%$ and a median $\chi^2_r$ of $0.96$, and only $0.01\%$ of it exceeds $\chi^2_r = 2$. We therefore use $\chi^2_r > 2$ to identify triggers that are inconsistent with being unlensed. These are $24\%$ of the injections, two thirds of them lying inside the strip, and their median loss is $37\%$. This implies that the penalty measured here is a response to the lensing distortion rather than to the discreteness of the bank.

This behavior can be understood from the interference between the two images, which modulates the amplification with a period $\fml = 1/\dt$ in frequency. We refer to one period of this modulation as a fringe. The $\chi^2$ bin boundaries are fixed by the unlensed template, so the fringes are sampled unevenly from bin to bin and the per-bin contributions depart from their expected values. 

The shaded strip in bottom-left panel of Fig.~\ref{fig:lens_plane_loss} marks where $\fml$ lies within the $20$--$1000$~Hz analysis band. We include it as a reference, since fringes outside the band cannot be sampled by the consistency test at all. However, we find that lying inside the strip is not sufficient to be discarded. Three effects control the size of the penalty: (i) the frequency at which the fringes fall, (ii) their depth, and (iii) the local width of the $\chi^2$ bins.

We first consider the frequency at which the fringes fall. The $\chi^2$ test responds only where the template carries power, and for the masses considered here most of the SNR is accumulated below $100$~Hz. Restricting to injections inside the strip with deep fringes, the median loss is $\sim9\%$ at $\fml \simeq 583$~Hz, $17\%$ at $185$~Hz, $41\%$ at $57$~Hz, and $37\%$ at $26$~Hz. This indicates that the upper edge of the strip overestimates the sensitive region by about an order of magnitude in $\fml$, and that the penalty peaks when the fringe spacing falls in the most sensitive part of the band. Since consecutive fringes are separated by $\fml$ in frequency, the number of them across an interval of width $W$ is $W/\fml = W\dt$. The bottom-right panel of Fig.~\ref{fig:lens_plane_loss} shows the three following regimes of interest. At $\dt = 0.3$~ms the whole band contains only a tenth of one oscillation, so $|\Ffac|$ acts as a slowly varying rescaling of the amplitude. At $\dt = 16$~ms the band contains about four oscillations while a single bin contains only $5\%$ of one, so neighbouring bins sample different parts of the pattern and their power departs from the expectation. At $\dt = 1$~s even the narrowest bin contains three oscillations, which average out within it.

We now consider the depth of the fringes, which is set by the relative magnification $\murel$. This quantity depends on $y$ alone, and is shown on the right-hand axis of the bottom-left panel of Fig.~\ref{fig:lens_plane_loss}. For injections inside the strip, the median loss rises from $3.4\%$ for $\murel < 0.2$ to $19.8\%$ for $\murel > 0.8$, and the median $\chi^2_r$ rises from $1.10$ to $1.83$ over the same range. This implies that both maps are organized by the same pair of variables, since $\murel$ is also the quantity that sets the depth of the fitting-factor loss in Eq.~\eqref{eq:ffloor}: $\dt$ determines where the loss occurs and $\murel$ determines how deep it goes.

We find that the relevant quantity is the local width of the $\chi^2$ frequency bins at the frequency where the fringes fall, rather than their width averaged over the band. For the non-spinning $(30, 30)\:\Msun$ template of Fig.~\ref{fig:lens_plane_loss}, they range from $3.3$~Hz near $40$~Hz to $18$~Hz near $200$~Hz, against a mean of $6.8$~Hz across the band. We note that the fringes are averaged away only once $\fml$ falls below the local bin width of a few hertz, rather than below the $20$~Hz lower edge of the band. Overall, the penalty is largest for $\dt$ between $10$ and $30$~ms ($\fml \simeq 58$~Hz), where the median loss is $33.4\%$ and the median $\chi^2_r$ is $2.80$, and it remains above $25\%$ out to $\dt \simeq 100$~ms. The penalty therefore peaks when the fringe spacing falls inside the most sensitive part of the detector band. It falls to $7.2\%$ for $\dt$ between $0.1$ and $0.3$~s, and to $2.4\%$ beyond $0.3$~s, so the suppression sets in once $\fml$ drops below a few hertz. This is expected, as the bins carry equal expected SNR$^2$, making them template-dependent, and the low-frequency bins are therefore only a few hertz wide even when the band-averaged width is an order of magnitude larger. This further modulates the suppression across the mass range since a $50$--$50\,\Msun$ template is tested with $20$ bins and a $10$--$10\,\Msun$ template with up to $73$.

At the opposite end, for $\dt$ below $3$~ms the median loss is $7.4\%$ with a median $\chi^2_r$ of $1.25$, which is above the control value of $0.6\%$ quoted earlier. We note that nearly all injections in this region have $w y^2 < 1$ at $100$~Hz and therefore lie outside the geometric-optics regime in which the two-image model of Sec.~\ref{sec:statistic} is derived, which explains the large effect due to lensing. However, we find in Sec.~\ref{sec:recovery} that the two-image fit removes this residual penalty as effectively as it removes the fringe penalty, reducing the median loss in reduced $\chi^2$ in this region.

We note that the fitting-factor map of Fig.~\ref{fig:ff} and the consistency-test map of Fig.~\ref{fig:lens_plane_loss} do not represent the same quantity, even though both are shown over the $(\ML, y)$ plane. At large $\dt$ the images are separated by more than the autocorrelation width of the template, so a single template matches one image cleanly and misses the other. The fitting factor then sits at the floor of Eq.~\eqref{eq:ffloor}, while the recovered waveform is a clean unlensed chirp and the consistency test is respected. At intermediate $\dt$, however, the images interfere within the band and the trigger is penalized even though a substantial fraction of the SNR can be collected. The two losses therefore peak in different regions of the lens parameter space, and the overall response of the search is their combination.

This result has a practical consequence for the design of a lensed search. Extending the template bank to contain lensed waveforms addresses Eq.~\eqref{eq:lossff} by supplying templates that match the distorted signal, but it does not address Eq.~\eqref{eq:lossveto}, which is a property of the statistic rather than of the bank, since a lensed template recovers the SNR but the $\chi^2$ is still measured against whichever single waveform was selected. The statistic constructed in the next section addresses both, without modifying the bank. It recomputes the $\chi^2$ under the lensing hypothesis, which removes Eq.~\eqref{eq:lossveto}. It also collects the matched-filter power at both image arrival times, which recovers most of Eq.~\eqref{eq:lossff}.


\section{Two-image recovery statistic and re-maximization over the bank}
\label{sec:statistic}

In this section, we construct a statistic that addresses the loss identified in Sec.~\ref{sec:chi2loss} without modifying the template bank. We first show that a microlensed signal in geometric optics lies in the two-dimensional subspace spanned by a template and a time-shifted copy of itself. Two of the three lensing parameters can then be maximized over analytically, and only the time delay requires a search. We then recompute the signal-consistency test under the lensing hypothesis and describe the choice of lag grid. Finally, we re-maximize over a neighbourhood of the bank, since the template recovered by the search need not be the best one for the two-image model.

\subsection{Two-dimensional subspace spanned by lensed signals}
\label{sec:subspace}

Substituting Eq.~\eqref{eq:Fgeo} into Eq.~\eqref{eq:amp},
\begin{equation}
  \tilde h_L(f) = a\, \tilde h(f) + b\, \tilde h(f)\, e^{-2\pi i f \dt} ,
  \label{eq:span}
\end{equation}
with $a = \sqrt{|\mu_+|}$ and $b = i\sqrt{|\mu_-|}$. The lensed waveform is a linear combination of the template and a time-shifted copy of itself. The coefficients enter linearly and can be maximized over analytically, and only $\dt$ is nonlinear and requires a search. Of the three lensing degrees of freedom, namely the relative magnification, the Morse phase and the time delay, two are therefore free. This is the observation on which the rest of the paper rests, and it is the reason the statistic does not inherit the computational cost usually associated with the addition of several degrees of freedom. We emphasize that in geometric optics this subspace is complete rather than approximate. The point-mass lens equation is quadratic and has exactly two solutions, so there is no third image for the model to miss, and the only limitation is the breakdown of geometric optics itself at low $w$.

This representation can also be used outside of the geometric optics regime. We evaluate the exact point-mass amplification factor across the prior and ask how much of the lensed waveform lies in the span of Eq.~\eqref{eq:span}, using the same code that analyses the injections and a template carrying the injected parameters. For delays between $0.08$~ms and $2$~s, the subspace recovers between $98.5\%$ and $100\%$ of the optimal SNR, against $75\%$ to $99\%$ for a single template. The two limits work for different reasons. When $w y^2$ is large the amplification is a sum of two discrete images, which is what Eq.~\eqref{eq:span} describes. When $w y^2$ is small it is a smooth modulation across the band with no fringe pattern, and two complex coefficients with a short lag absorb it; the fitted lag is then comparable to the template autocorrelation width rather than to any physical delay, so the model acts as a generalized amplitude and phase correction. The geometric derivation therefore understates where the statistic works. This argument rests on the point-mass lens having exactly two images, and a lens producing more would need a wider family.

Note also that $a$ and $b$ are properties of the lens, not of the detector, since $\Ffac$ multiplies the whole strain. The ratio $b/a$ and the delay $\dt$ are therefore common to every detector in the network, and we exploit the delay by maximizing over a single $\dt$ shared across detectors.

\subsection{Construction}
Let $z(t)$ be the complex matched-filter output of the normalized template
$\hat h$ against the data, and
\begin{equation}
  R(\tau) = \langle \hat h , \hat h_\tau \rangle
\end{equation}
the template autocorrelation, where $\hat h_\tau$ is $\hat h$ delayed by $\tau$ and $\langle\cdot,\cdot\rangle$ is the noise-weighted inner product. Because $\hat h_\tau$ placed at time $t$ is $\hat h$ placed at $t+\tau$, the two filter outputs required are the \emph{same} series evaluated at two lags, and no additional filtering is needed, as also noted in~\cite{Lo:2024wqm}. Writing $\bm{z}_\tau = [\,z(t),\, z(t+\tau)\,]^{\rm T}$ and
\begin{equation}
  \bm{G}(\tau) =
  \begin{pmatrix}
    1 & R(\tau) \\ R^{*}(\tau) & 1
  \end{pmatrix} ,
  \label{eq:gram}
\end{equation}
the SNR maximized over $a$ and $b$ is the norm of the projection of the data onto the subspace,
\begin{equation}
  \rhoL^2(t,\tau) = \bm{z}_\tau^{\dagger}\, \bm{G}^{-1}(\tau)\, \bm{z}_\tau ,
  \label{eq:rhoL}
\end{equation}
with maximum-likelihood coefficients $[\hat a, \hat b]^{\rm T} = \bm{G}^{-1} \bm{z}_\tau$. This analytic maximization over coefficients that enter the waveform linearly is a similar step to that in the $\mathcal{F}$-statistic~\cite{Jaranowski:1998qm}. We evaluate Eq.~\eqref{eq:rhoL} at the trigger time from the unlensed search and maximize it over a grid in $\tau$, denoting the best-fit lag $\hat{\tau}$.

The statistic behaves correctly in both limits. As $\tau \to \infty$,
$R \to 0$ and $\bm{G} \to \bm{I}$, so $\rhoL^2 \to |z(t)|^2 + |z(t+\tau)|^2$; the two images are thus added in quadrature. As $\tau \to 0$ the two basis vectors coincide, $\bm{G}$ becomes singular, and the statistic reduces to $|z(t)|^2$. We exclude lags with $|R(\tau)| > 0.9$, which is both numerically necessary and the physical statement that images closer together than the autocorrelation width are not resolvable~\cite{Lo:2024wqm}.

We note that Eq.~\eqref{eq:rhoL} is a projection, so $\rhoL$ cannot exceed the optimal SNR of the data. This provides a useful check on an implementation, since the ordering of $R$ and $R^{*}$ in Eq.~\eqref{eq:gram} follows the convention used for $z$, and the opposite ordering returns values above the optimal SNR wherever $|R|$ is large.

We validated this construction against the full lensed waveform, independent of noise. For 99.9\% of the cases, the match is above 0.99; see Appendix~\ref{app:validation}.

\subsection{The consistency test under the lensing hypothesis}

By obtaining the shared delay $\hat\tau$ and the coefficients $(\hat a_i, \hat b_i)$ in each detector $i$, we recompute the power $\chi^2$ with the expected per-bin contributions evaluated from the fitted two-image model $\hat a_i \hat h + \hat b_i \hat h_{\hat\tau}$ rather than from the unlensed template. A microlensed signal is then consistent by construction, while an instrumental transient remains inconsistent, since it does not fit  a two-image model either. We denote the resulting reduced statistic $\chisqL$, with $2p-4$ degrees of freedom, and the associated re-weighted SNR $\newsnr_L$.

We note that this test is computed per detector and then pooled, similarly to what is done in the unlensed case. The detectors are combined coherently in the delay, since we maximize the summed $\rhoL^2$ over a single $\dt$ common to the network. Because $\Ffac$ multiplies the whole strain, the antenna pattern scales both images equally, so the ratio $\hat b_i / \hat a_i$ is common to the network even though the overall amplitude is not. A microlensed signal must therefore return the same ratio in every detector, while an instrumental transient has no such requirement. Our background is Gaussian, however, and further studies on real data are needed to understand the gain against instrumental transients.

\subsection{Choice of lag grid}
\label{sec:laggrid}

The grid in $\tau$ has to satisfy three requirements, and they are set by different scales. The spacing must be below about half the template autocorrelation width, which is of order a millisecond. With a coarser spacing the fitted model is offset from the one in the data, and the residual $\chi^2$ may remain large even for a signal that has been correctly identified. The extent must cover the prior in $\dt$, which reaches $9.05$~s at $\ML = 10^5\,\Msun$ and $y = 2$\footnote{For most GW signals, this would correspond to distinguishable images contained within one frame and, therefore, such cases also show how our method can naturally be extended to the case of strong lensing searches, even if we do not focus on this particular case in this work.}. If the grid does not reach far enough, the fit pins at the edge and the corresponding improvement is an artifact. The grid must also reach backwards. We evaluate the statistic at the trigger sample, which is the loudest sample within $\pm 0.1$~s of the arrival time of the signal in each detector. When the two images have nearly the same magnification, the louder of the two is sometimes the second one, and the first image then lies before the trigger sample. With a forward-only grid the fit cannot reach it, and the consistency penalty remains. We use a spacing of $0.4$~ms from $-1$~s to $+10$~s, which is $2.75\times10^{4}$ lags. The backward reach only has to cover the window in which the trigger sample is chosen, and $-1$~s leaves a margin of ten. Because Eq.~\eqref{eq:rhoL} is evaluated from two samples of an existing series, the cost of the grid is negligible, and the consistency test dominates the cost of the statistic.

The grid covers two different regimes. When the delay is short, the two images overlap in the band and interfere, and the consistency test removes SNR from a signal that the search has already found. The statistic recovers that SNR. When the delay is longer than about a second, the two images are well separated, and the search finds the brighter one as a clean unlensed chirp. We measure a median $\chi^2_r$ of $1.0$ and a median loss of $1.4\%$ in this region. The brighter image is therefore detected on its own, regardless of whether the fainter one is found.

The loss that remains at long delay is the SNR of the second image, which is $8.8\%$ in the median. Here the statistic adds the second image in quadrature. This is a joint detection of both images, and not the recovery of a signal that would otherwise be missed, though this may enable to recover signals which may not even be found with dedicated approaches~\cite{McIsaac:2019use, Li:2023zdl}. 

\subsection{Re-maximization over the bank}
\label{sec:rescan}

The search selected its template to maximize the \emph{one-image} SNR against a distorted signal. The template is therefore displaced from the true parameters in a way that depends on the lens. The best template for the two-image model need not be the same one. We therefore also maximize over a neighbourhood of the recovered template.

A fixed number of neighbours is not a suitable choice, since the bank is far denser at low chirp mass than at high. The nearest $200$ templates span $0.09\%$ in chirp mass at $\Mc = 8\,\Msun$ and $19\%$ at $\Mc = 70\,\Msun$. We therefore select neighbouring templates within a fractional chirp-mass window around the template recovered by the search,
\begin{equation}
  \frac{|\Delta \Mc|}{\Mc} \le k\, \frac{\Mc}{\rho} ,
  \label{eq:rescanwindow}
\end{equation}
where $\Mc$ is the chirp mass of the recovered template and $\rho$ is the recovered network SNR. The window is motivated by the displacement between the recovered and injected chirp masses measured in our simulations, which increases with chirp mass, since a light system has more cycles in band and is pinned down more tightly in fractional terms, and decreases with SNR. The $90^{\rm th}$ percentile of that displacement follows Eq.~\eqref{eq:rescanwindow} with $k$ between $0.09$ and $0.13$, over a factor of $35$ in $\Mc/\rho$. We use $k = 0.20$, for which the window contains the true chirp mass of the injection in $96.7\%$ of cases, and in $89.0\%$ of those in the high-$\ML$, low-$y$ corner. The window size becomes very wide for a heavy trigger with low SNR. At $\Mc = 43\,\Msun$ and $\rho = 12$ it gives a window of $72\%$ chirp mass. A wider window adds templates that are far from the signal, which raises the trials factor without improving the fit. We therefore cap the window at $|\Delta \Mc|/\Mc = 0.5$, which still contains the true chirp mass for $99\%$ of the injections.

We first rank them by $\rhoL$ alone, which costs about a tenth of the full statistic per candidate. We then compute the full statistic for the best $20$ and select on $\newsnr_L$. Since the initially recovered template is always included in the process, the re-maximization cannot lead to a worse result than the initial one.

\section{Recovery}
\label{sec:recovery}

\begin{figure*}[!tp]
\includegraphics[width=0.99\textwidth]{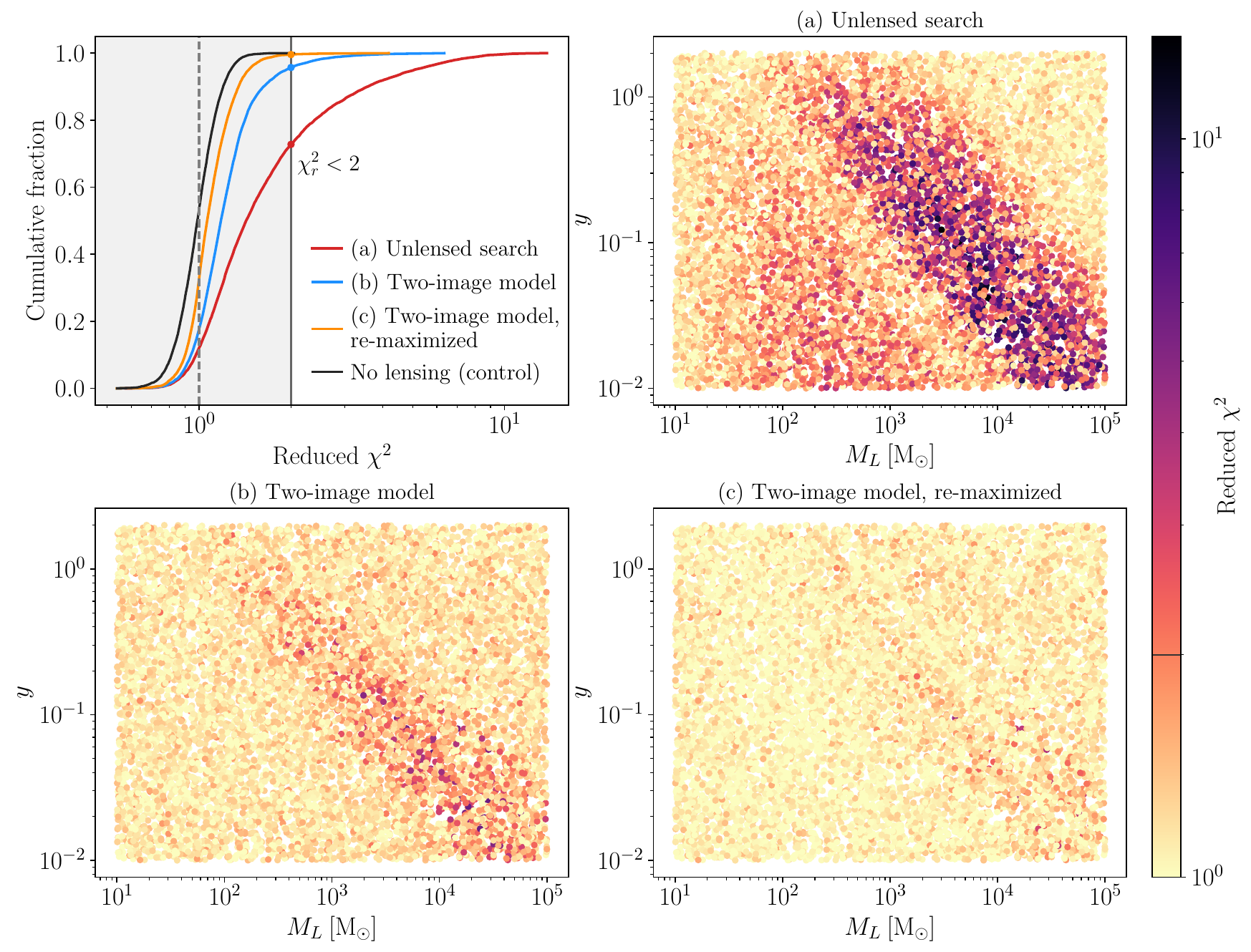}
  \caption{Network reduced $\chi^2$ for the microlensed injections of Fig.~\ref{fig:lens_plane_loss}, for the three analyses considered in this work. The first panel gives the cumulative distribution for the unlensed search (red), the two-image model (blue) and the two-image model re-maximized over the bank (orange), with the unlensed control in black. The remaining panels map the same quantity over the lens parameter space, for (a) the unlensed search, (b) the two-image model and (c) the two-image model re-maximized. The injections and the templates used for the search are identical in the three maps; only the model the residual is measured against differs. The diagonal band present in (a) is weaker in (b) and almost gone in (c).}
  \label{fig:recovery_chisq}
  \vspace{0.1cm}
  \includegraphics[width=0.99\textwidth]{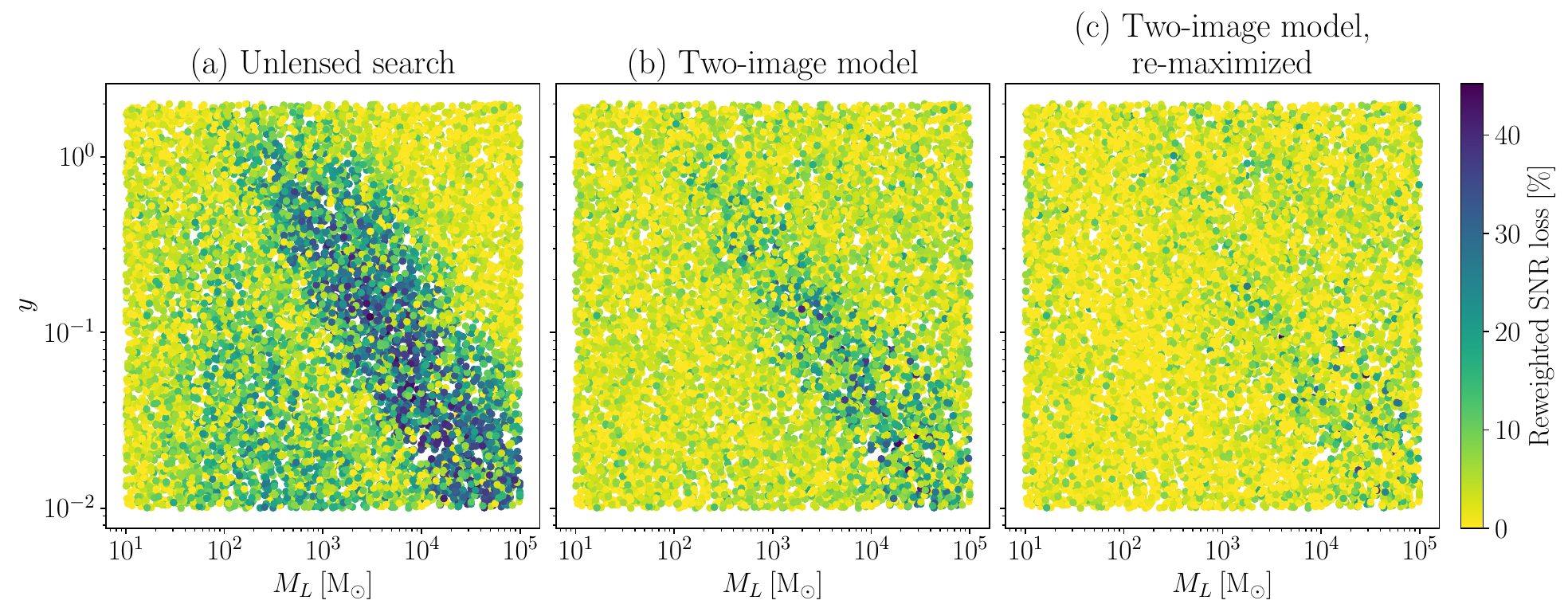}
  \caption{Same as Fig.~\ref{fig:recovery_chisq}, but for the fraction of the network SNR removed by the re-weighting.}
  \label{fig:recovery_loss}
\end{figure*}

In this section, we apply the statistic and methods outlined in Sec.~\ref{sec:statistic} to the same injections as Sec.~\ref{sec:selection}. We compare the performance of three analyses of the same microlensed dataset:

\begin{enumerate}
  \item[(a)] \emph{Unlensed search:} The standard \pycbc\ search, which filters the lensed signal against the unlensed template bank and applies the usual signal-consistency test. Nothing in it is adapted to lensing, and it is the baseline against which the other two are measured. This is the same as in Sec.~\ref{sec:selection}.
  \item[(b)] \emph{Two-image model:} The two-image fit of Sec.~\ref{sec:statistic}, built on the template recovered by the search. 
  \item[(c)] \emph{Two-image model, re-maximized:} The same fit, after re-maximizing over the neighborhood of Eq.~\eqref{eq:rescanwindow}. This differs from (b) only in that the data are filtered again against each neighboring template, so it is the more expensive of the two.
\end{enumerate}

We also run the two-image fit on the lensed injections using a template with the injection's own intrinsic and lensed parameters, instead of one drawn from the bank. A search cannot do this, and we include it only to bound how much the choice of template matters.

We then repeat the whole analysis on a separate control set of \emph{unlensed} injections. These have the same masses, spins, sky positions, orientations and distances as the lensed set, and we add them to the same data. Since the distances were set so that the lensed signal reaches its target SNR, removing the lensing leaves the control quieter, with a median recovered SNR of $13.4$ against $26.0$. The two-image model has two free parameters more than a single template. It therefore absorbs some ordinary waveform mismatch even when no second image is present, and the control measures how much. Any improvement seen in the control is not caused by lensing.

The numbers in this section come from the follow-up rather than from the search. The search lets each detector recover its own template, while the follow-up uses one template in both, taken from the louder detector. The quieter detector is then filtered with a template it did not itself recover, which raises its $\chi^2$ a little. The median loss for (a) is therefore $10.5\%$ in this section against $9.2\%$ in Sec.~\ref{sec:chi2loss}. The network SNR is the same in both, so the difference lies in the consistency test and not in what the search recovered.

\subsection{Removal of the consistency penalty}

Figs.~\ref{fig:recovery_chisq} and \ref{fig:recovery_loss} compare the network reduced $\chi^2$ and the fractional loss in $\newsnr$ for analyses (a) and (b). The diagonal band of Sec.~\ref{sec:selection} is clearly present in (a), while it is reduced for (b), when the model includes the lensing statistic.

When we perform the two-image search on the lensed injection campaign, the median $\chi^2_r$ falls from $1.43$ to $1.18$ and the median loss from $10.5\%$ to $4.5\%$. Inside the strip the fall is larger, from $1.73$ to $1.25$ and from $17.2\%$ to $6.3\%$. Re-maximizing over the bank (analysis (c)) further reduces the median $\chi_{r}^2$ to $1.06$ and the median loss to $1.6\%$, and inside the strip to $1.07$ and $1.9\%$. The unlensed control has a median $\chi^2_r$ of $0.99$ under (c). Therefore, the re-maximization enables one to recover performances similar to the one seen for unlensed data with an unlensed search. 

The residual that remains tells us something about the model. A deficiency of the two-image model would depend on the lens parameters, and the band would still be visible. Binned in $\dt$, the median $\chi^2_r$ under (c) varies only between $1.02$ and $1.13$, against $1.12$ to $1.37$ under (b), so the band is essentially gone after re-maximization. What is left is the mismatch floor of the bank, the fluctuation of $\chisq$ about unity, and the cost of the two extra degrees of freedom, none of which depend on the lens. The control sets the scale of what remains. Since $\chisq$ grows with $\rho^2$ at fixed fractional mismatch, we compare the two sets in the same range of recovered SNR, and for $\rhoL$ between $20$ and $40$ the lensed median and the control median differ by less than $0.02$. When we perform the two-image searches ((b) and (c)) on the unlensed injections, the median $\chi^2_r$ rises only from $1.023$ to $1.056$ and from $0.986$ to $0.989$. The two-image model therefore describes these signals completely, and what is left after the fit is the ordinary floor of the search rather than anything due to the lens model.

\begin{figure}[t]
  \centering%
    \includegraphics[width=0.47\textwidth]{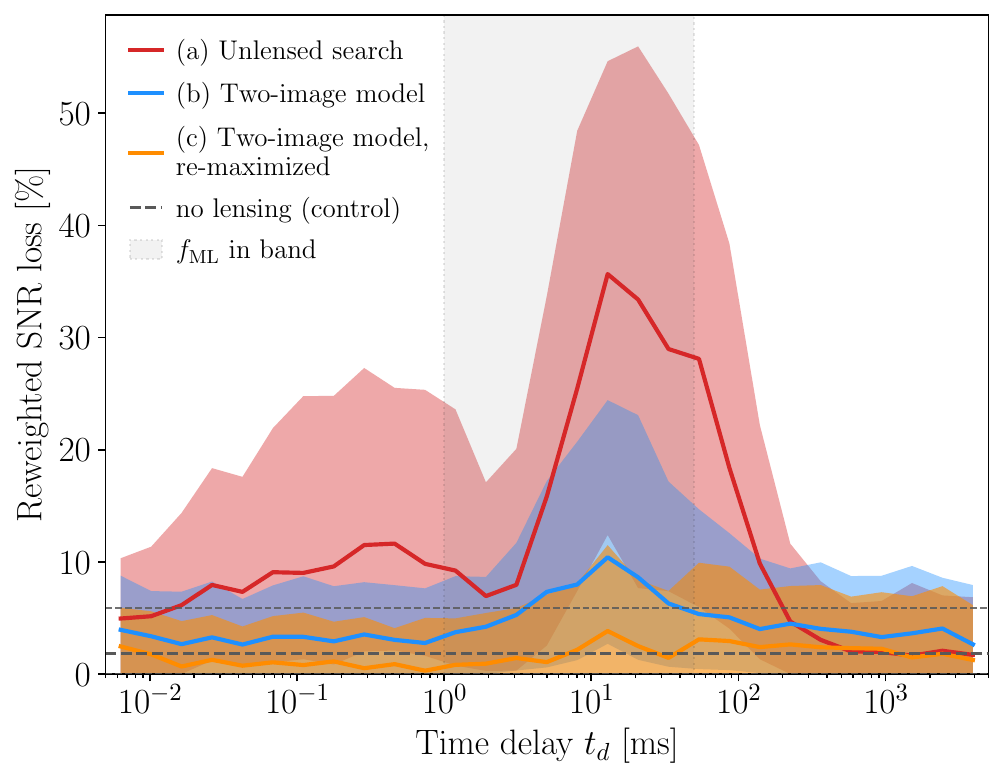}
  \caption{Fraction of the network re-weighted SNR removed by the consistency test as a function of the image time delay, for the unlensed search (red), after the two-image fit (blue), and after re-maximizing over the bank (orange). Solid  lines are medians in logarithmic bins of $\dt$; shaded regions span $16$--$84$\%, and the dashed lines give the median and the $16$--$84$\% range of the unlensed control, reweighted to the recovered-SNR distribution of the lensed set. The grey strip marks where the fringe frequency $\fml = 1/\dt$ lies within the $20$--$1000$~Hz band. The penalty peaks at $\dt \simeq 12.9$~ms, where the median loss falls from $35.7$\% to $10.4$\% and to $3.8$\%; inside the strip it falls from $17.2$\% to $6.3$\% and to $1.9$\%. The response extends beyond the strip because the fringes are averaged away only once $\fml$ drops below the width of a $\chisq$ bin, $2$--$4$~Hz at low frequency, rather than below the edge of the band.}
  \label{fig:loss_vs_td}
\end{figure}

\subsection{Dependence on the image time delay}
\label{sec:delay}

Fig.~\ref{fig:loss_vs_td} shows the fractional loss in $\newsnr$ as a function of the image time delay, for (a), (b) and (c). The shaded regions span the $16^{\rm th}$ to $84^{\rm th}$ percentiles, and the dashed lines give the unlensed control. The grey strip marks where $\fml$ lies within the $20$--$1000$~Hz band. We find that the gain is not uniform in $\dt$, and we describe the three regimes below.

The penalty is largest at $\dt = 12.9$~ms. There the median loss falls from $35.7\%$ in (a) to $10.4\%$ in (b) and $3.8\%$ in (c). Across the strip it falls from $17.2\%$ to $6.3\%$ and to $1.9\%$. This is the regime where the two images interfere inside the band. The consistency test responds to the distortion, and the statistic removes it.

At short delay the two images are not separable by the matched filter. These configurations lie in the wave-optics regime of Sec.~\ref{sec:chi2loss}, where the amplification is a smooth modulation across the band and not a fringe pattern. Below $1$~ms the median loss still falls from $9.0\%$ to $3.1\%$ in (b) and $0.8\%$ in (c). Although the absolute improvement is smaller than inside the strip, the remaining loss is also smaller, showing that the two-image model is effective even in this regime.

At long delays, beyond $0.3$~s, the behaviour changes again. The median loss in (a) is only $2.2\%$. In (b) it is $3.6\%$ and in (c) $2.1\%$, so the fit costs a small amount here instead of gaining. This is the price of the two extra degrees of freedom when there is no distortion to remove. At long delay the two images are well separated, so the statistic adds them in quadrature. The median $\rhoL/\rho$ is $1.13$, which means the fainter image carries about half the SNR of the brighter one. As discussed in Sec.~\ref{sec:laggrid}, the fainter image may be below the detection threshold on its own, but its contribution can still improve the joint detection. Thus, the gain in this regime comes from combining both images rather than reducing the consistency-test penalty.

\begin{table}
  \begin{tabular}{lcccc}
    \hline\hline
    template & $\chisqL$ & loss & $\rhoL$ & $\newsnr_L$ \\
    \hline
    (b) recovered      & $1.18$ & $4.5\%$ & $28.1$ & $26.0$ \\
    (c) re-maximized   & $1.06$ & $1.6\%$ & $28.4$ & $27.4$ \\
    true parameters    & $1.09$ & $2.3\%$ & $28.4$ & $26.7$ \\
    \hline\hline
  \end{tabular}
  \caption{The two-image fit built on three different templates. For each quantity, we report the median over the full set. The last row uses the injection's own parameters, which a search cannot access.}
  \label{tab:template}
\end{table}

Table~\ref{tab:template} compares (b), (c) and the fit built on the injection's own parameters. We find that $\rhoL$ is mostly insensitive to the template, while $\newsnr_L$ is not. The median $\rhoL$ varies by $1.2\%$ between the three, the median $\newsnr_L$ by $5.6\%$, and the median $\chisqL$ by $11\%$. The reason is that $\chisqL$ responds sharply when the two-image fit is built on a mismatched template, and $\newsnr_L$
inherits that response.

We also find that re-maximization exceeds the fit built on the injection's own parameters. The median $\newsnr_L$ is $27.4$ after re-maximization and $26.7$ with the true parameters. This is expected, since the re-maximization can pick a template that suits the particular noise realization, while the injection's parameters are fixed. The gap is $2.8\%$, and it measures how much the re-maximization takes from the noise. Using an unlensed background, as shown in Sec.~\ref{sec:background}, corrects for this.

\begin{figure}
    \centering
    \includegraphics[width=\linewidth]{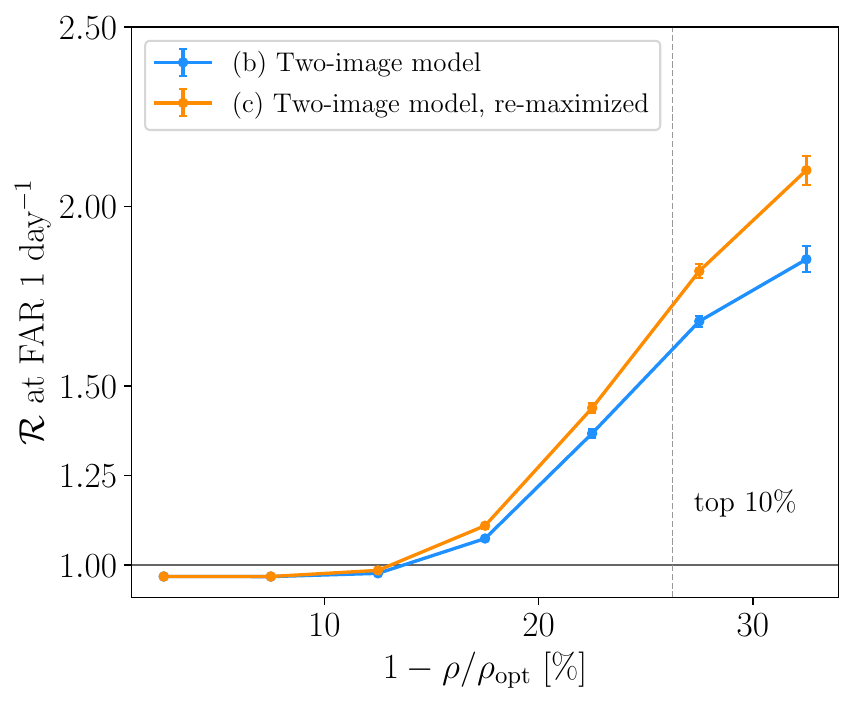}
    \caption{Relative detection volume $\mathcal{R}$, Eq.~\eqref{eq:volratio}, against the fitting-factor loss, Eq.~\eqref{eq:lossff}, in differential bins. Blue combines the unlensed search with the two-image model (b), orange with the re-maximized version (c). The dashed line marks the decile with the largest loss, above $26\%$, where $\mathcal{R}$ is $1.75$ and $1.93$. Error bars are bootstrapped over the injections. The same statistic applied to the unlensed control gives $0.97$.}
    \label{fig:gain}
\end{figure}

\section{Search Sensitivity}
\label{sec:background}

A lower $\chi^2$ does not by itself mean a better search, because $\rhoL$ is maximized over a larger space than $\rho$ and its background is higher. Therefore, we use a background on which we run the same analysis to obtain a distribution of the statistic under the null hypothesis. So, we apply the same analysis with the same settings, to stretches of the same data that contain no injection. This includes the trials factor by construction, and needs no analytic counting of degrees of freedom.

We use 0.92 days of injection-free coincident analysis time. The data are taken in separate stretches, so a slide that shifts one end past a stretch boundary contributes no background. We form coincidences in the usual way~\cite{Davies:2020tsx}, and repeat them with one detector shifted by multiples of $15$~s in both directions. This gives $100$ slides and $83.3$~days of background. We cluster the coincidences over $10$~s, treating each slide separately so that events from different slides cannot merge. We then run the same follow-up on the survivors that we run on the injections, including the re-maximization over the bank. Every background event therefore carries the same two numbers an injection carries: $\newsnr$ from the single recovered template, and $\newsnr_L$ from the two-image fit. This leaves $1.9\times10^{4}$ independent background events.

A threshold is then a matter of counting. If we allow one false alarm per day, and the background is $83.3$~days long, we are allowed $83$ background events above the threshold. We sort the background by $\newsnr$ and read off the $83$rd loudest value, which gives $t_0 = 8.29$.

We do not replace the unlensed search with the two-image one. Most signals are not lensed, and for them the two-image fit gains nothing while its threshold is higher, so a search ranked on $\rhoL$ alone would be worse than the unlensed search almost all of the time. We instead keep a candidate when \emph{either} statistic is loud enough, and we call this the combined search. A candidate counts as a false alarm if either statistic crosses its threshold, so the two share one budget and each has to cut higher up its own sorted list. Requiring that the rate at which either fires equals one per day gives $t_a = 8.38$ on $\newsnr$ and $t_L = 9.87$ on $\newsnr_L$. The rise from $t_0$ to $t_a$ is what the unlensed search pays for the second statistic. The larger value of $t_L$ follows from the lag grid and the bank, over which $\rhoL$ is maximized: a search over a larger space has a louder background and must take a higher threshold at fixed false-alarm level~\cite{Capano:2016dsf, Narola:2022aob}. We also measure the background for (b) separately, without the re-maximization, and find $t_L = 9.85$. The trials added by searching the bank are therefore small.

We next estimate how far each search can see. Moving a signal further away lowers the SNR as the inverse distance. It also lowers the consistency penalty, since the $\chisq$ excess is a fixed fraction of the signal power and therefore grows as $\rho^2$. The second effect matters most for the heavily penalized signals. Neglecting it would understate their reach by the re-weighting factor of Eq.~\eqref{eq:newsnr}, and the volume by its cube, which is $1.4$ at the median $\chi^2_r = 1.43$ of our injections and $22$ at $\chi^2_r = 10$. We therefore keep both effects. A signal stays above a threshold $t$ until it is moved $x$ times further away, where $x$ solves
\begin{equation}
  \newsnr(x) = t ,
  \qquad \text{with } \ \rho \to \frac{\rho}{x}
  \ \text{ and } \ \chi^2_r \to 1 + \frac{\chi^2_r - 1}{x^{2}}
  \ \text{ in Eq.~\eqref{eq:newsnr}} ,
  \label{eq:reach_a}
\end{equation}
and $\rho$ and $\chi^2_r$ are the values measured at the distance at which the signal was injected. We solve this numerically for each injection and call $x$ the reach, in units of that distance. We write $x_0$ for the reach of the unlensed search, obtained with $t = t_0$.

The combined search keeps the candidate when either statistic crosses its threshold, so it still finds the signal as long as one of them does. Its reach is therefore the larger of the two,
\begin{equation}
  x = \max \left( x_a , \ x_L \right) ,
  \label{eq:reach_c}
\end{equation}
where $x_a$ and $x_L$ are obtained from Eq.~\eqref{eq:reach_a} with $(\rho, \chi^2_r, t_a)$ and with $(\rho_L, \chisqL, t_L)$ respectively.

A source can be seen anywhere inside a sphere of that reach, so the volume it represents goes as the cube. Adding these over the injections gives the
relative detection volume,
\begin{equation}
  \mathcal{R} = \frac{\sum_i x_i^{3}}{\sum_i x_{0,i}^{3}} ,
  \label{eq:volratio}
\end{equation}
where the sums run over the injections. We add the cubes rather than averaging the ratios, because a volume is a sum of volumes. The same construction is used
in Ref.~\cite{Mozzon:2020gwa} to compare two ranking statistics at fixed false-alarm rate, where it is checked against a count of found injections. This is not the sensitive volume of the search, which would be obtained by counting found injections drawn uniformly in volume; it is the volume implied by the reach of each search on the same set of signals. We do not count found and missed injections here, because the distances are set so that the optimal network SNR is uniform in $[8,50]$: the unlensed search recovers $96.9\%$ of them at their own distance, and a count would have almost nothing to measure.

Fig.~\ref{fig:gain} shows the relative detection volume against the fitting-factor loss, Eq.~\eqref{eq:lossff}, which measures how much of the signal the unlensed bank failed to describe. Unlike $\chi^2_r$, it does not depend on the distance at which the signal was placed. Combining the unlensed search with (b) gives $\mathcal{R} = 1.04$ over the full set, and combining it with (c) gives $1.06$.\footnote{At one per $50$ days the gain is $1.10$ for (b) and $1.12$ for (c). It is larger there because $t_a = t_0$ at that rate, so the unlensed search pays nothing for the second statistic.} For the $10\%$ of injections with the largest loss, above $26\%$, the values are $1.75$ and $1.93$, and inside the strip they are $1.08$ and $1.12$. When ordered instead by the consistency statistic, triggers with $\chi^2_r > 2$ give $1.12$ and $1.16$; these are $27\%$ of all the injections. Applied to the unlensed control the same statistic gives $0.97$, as expected for signals with no second image to recover. The gain is nearly independent of the injected SNR, varying by less than $0.04$ over $\rho_{\rm opt} \in [8,50]$, because the reach of Eq.~\eqref{eq:reach_a} does not depend on how loud the injection was placed. We note that $\mathcal{R}$ is a ratio measured at a fixed lens-parameter prior. Translating it into an expected increase in the astrophysical detection rate would also require the distribution of $(\ML, y)$ in the Universe and the corresponding optical depth.

The gain is also not spread evenly over the injections. The two-image threshold is higher than the one-image one by $18\%$, so the fit only pays when it reaches further in spite of that. This happens for $28\%$ of the injections under (b) and $33\%$ under (c), and for those the relative detection volume is $1.43$ and $1.49$. For the rest the combined search costs $3\%$, which is the threshold the unlensed search gave up. The quoted gain is the average of the two.

\begin{figure*}
  \includegraphics[width=\textwidth]{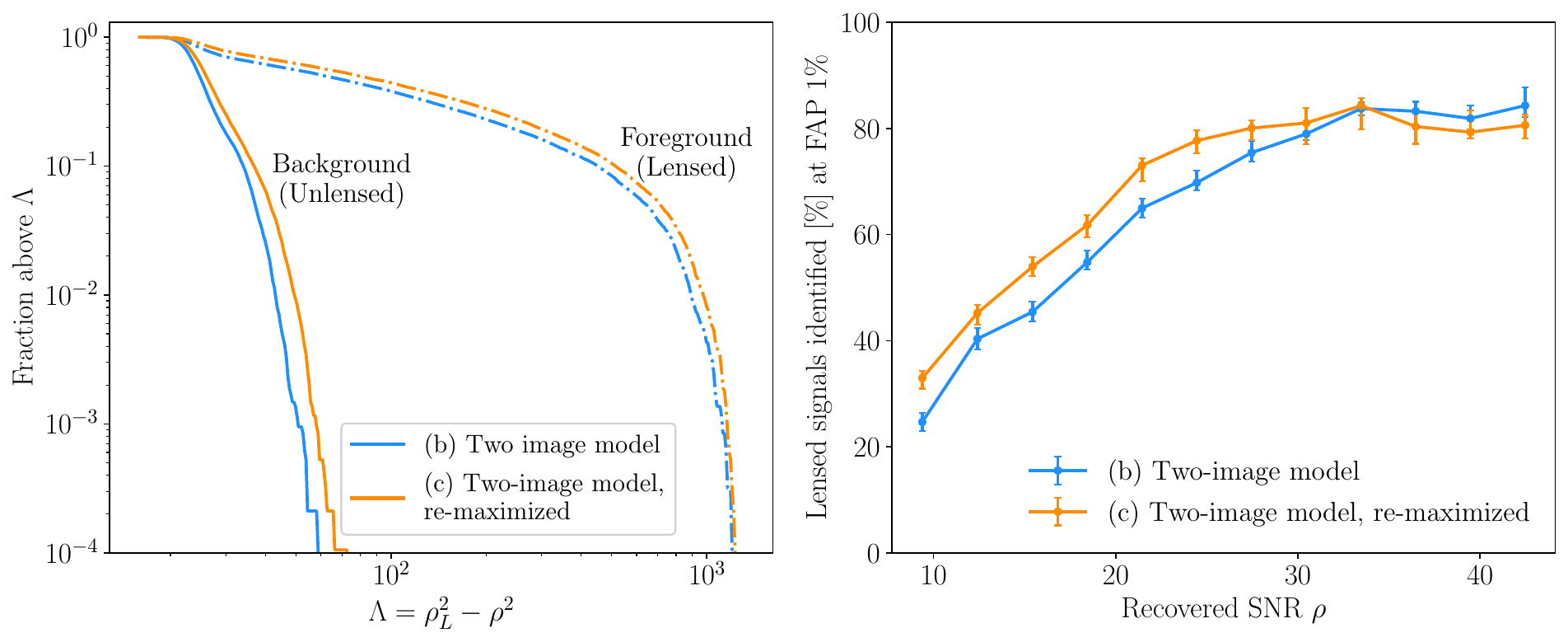}
  \caption{Identification of lensed signals using $\Lambda = \rho_L^2-\rho^2$, twice the difference in maximized log likelihood between the two-image and the one-image model. The left panel shows the fraction of injections above a given $\Lambda$, for the unlensed control set (solid) and for the lensed set (dash-dotted), where colour denotes the choice of template. The unlensed curve gives the false-alarm probability of the identification. The right panel shows the fraction of lensed signals identified at a 1\% false-alarm probability, with the threshold obtained from the control set separately in each bin of recovered SNR. Error bars are obtained by bootstrapping the control and the lensed sets jointly. The efficiency increases with SNR, and is higher when the template is re-maximized over the bank.}
  \label{fig:fap_comparison}
\end{figure*}

\section{Identifying lensed signals}
\label{sec:identification}

In this section, we investigate a second question not addressed by our method so far: given that our adapted search recovers a trigger, is it genuinely lensed? This is distinct from the question of Sec.~\ref{sec:background}, which distinguishes between astrophysical signal and noise. It requires a different null background. An unlensed signal is not noise, and the two-image model can fit such a signal, so the noise background is not sufficient to establish the lensed nature of the detection. The appropriate null is a population of unlensed signals, for which we use the control set of Sec.~\ref{sec:recovery}.

We characterize the two hypotheses using
\begin{equation}
  \Lambda = \rho_L^2 - \rho^2 ,
  \label{eq:lambda}
\end{equation}
which is twice the difference in maximized log likelihood between the two-image and the one-image model, and therefore the likelihood ratio between them, maximized rather than marginalized over the lens parameters. A Bayes factor would instead marginalize over the relative magnification, the time delay and the Morse phase under a prior, and would pay an Occam factor for that additional freedom. Here we measure the corresponding penalty directly, since whatever the extra freedom gains on a signal that is not lensed is given by the unlensed distribution in Fig.~\ref{fig:fap_comparison}, from which we read the threshold.

The statistic has three lensing degrees of freedom: the relative magnification, the Morse phase, and the time delay. The first two enter linearly and are maximized over analytically, and at a fixed delay, $\Lambda$ would then follow a $\chi^2$ with two degrees of freedom and a median of $1.4$. The third is searched, and that is where the penalty comes from: the null is not a typical draw but the largest of the $2.75\times10^4$ lags of the grid of Sec.~\ref{sec:laggrid}. We do not need the analytic form of that distribution, since the significance is derived directly from analyzing the null background.

The median of the null distribution is nearly independent of the recovered SNR. However, maximization over the delay and the template bank introduces additional trials~\cite{Gross:2010qma, Tenorio:2021wad}. We find that this affects the tail of the distribution: the $99^{\rm th}$ percentile of $\Lambda$ rises from $35$ at $\rho \simeq 20$ to $54$ at $\rho \simeq 40$ when we re-maximize over the bank, while it stays between $33$ and $39$ when we do not. For a fixed fractional waveform mismatch, the residual power grows with $\rho^2$, so a louder signal gives the re-maximization more residual power to absorb. A neighbouring template may then better fit this residual or a noise fluctuation resembling a second image. We therefore set the threshold separately in each bin of recovered network SNR, so that the false-alarm probability is fixed for a given recovered SNR. The resulting efficiencies are shown in the right panel of Fig.~\ref{fig:fap_comparison}. Using the recovered rather than the injected SNR ensures that the lensed injections are compared against the unlensed control at similar recovered SNR, despite their different overall SNR distributions.

At a false-alarm probability of $1\%$, we recover $71\%$ of the lensed injections with $10 < \rho < 45$, increasing from $45\%$ at $\rho \simeq 12$ to $82\%$ at $\rho \simeq 36$. This averaged number reflects our lens prior rather than the statistic itself, since half of the injections in this prior are only weakly lensed. We therefore also report the efficiency conditioned on the reduced $\chi^2$ of the template recovered by the search. Conditioned in this way, the efficiency is $99.4\%$ for $\chi^2_r > 2$ and above $99.8\%$ beyond $3$. This indicates that once the lensing is visible to the search, the identification is essentially complete. An exact comparison with results from a dedicated parameter estimation approach, such as \gravelamps~\cite{Wright:2021cbn}, is left for future work.

We also note two further features. First, $54\%$ of our injections have $\chi^2_r < 1.5$, where the search itself barely responds to the lensing, and the statistic still identifies $52\%$ of them. This shows that $\Lambda$ responds to second images even when it is relatively weak. Second, the identified events are not marginal. For $\chi^2_r > 2$, $98.3\%$ of them are louder than every one of the $\sim 10^4$ unlensed injections in the control set, so their false-alarm probability lies below the smallest value this control can resolve, which is $\sim 10^{-4}$. Quoting a smaller value would require a larger control set rather than a different statistic. Such a large set, also needed to assess the significance of triggers in usual searches~\cite{LIGOScientific:2021izm, LIGOScientific:2023bwz, LIGOScientific:2025cwb}, would be easier to obtain with this approach since it does not require to run parameter estimation.

Finally, we observe that the identification efficiency behaves differently from the gain in recovered SNR of Sec.~\ref{sec:delay}. The latter is concentrated in the band of delays where the consistency test is most effective, whereas the identification efficiency stays above $90\%$ from $5$~ms out to one second. These two results are consistent with each other. At long delays the search already recovers the signal without assistance, so there is no SNR to regain, while the second image remains present in the data and can still be identified in the second step.

Two caveats apply to these efficiencies. First, we generate and analyze the control set with the same waveform model, so waveform systematics are absent by construction. With a different model, the two extra degrees of freedom would absorb part of the intrinsic mismatch. This raises $\Lambda$ in the control, and therefore raises the threshold and lowers the efficiency. The residual power available to be absorbed is $\rho^2\epsilon$ for a model mismatch $\epsilon$. For the aligned-spin binaries considered here, this is small compared with the thresholds quoted above. The same limitation applies to the standard search, whose templates also omit higher-order modes, precession and eccentricity, and whose consistency test responds to any of them. The difference is that such a deviation costs the standard search only SNR, while it may be identified as lensed by $\Lambda$. The fitted delay is then comparable to the template autocorrelation width, where the two-image model acts as a generalized amplitude and phase correction, so this is the regime in which the identification is least reliable. Second, the noise is Gaussian, so a transient overlapping a signal cannot raise $\Lambda$ here. In real data, the control would be injections into the same data, as is done for background estimation in usual searches. The lens consistency requirement discussed below would then provide a further discriminant.

Two further extensions concern the identification presented here. The statistic maximizes over the lens parameters, and a population-weighted Bayes factor would marginalize over them instead, at the cost of a prior. The fit also returns the relative magnification, the time delay and the Morse phase, which we do not use beyond ranking; when the images are separable these are measurements of the lens, and comparing them against the point-mass relation is a consistency test that an instrumental transient has no reason to pass. Explicitly extending our work in that direction is left for future work.

\section{Discussion and conclusions}
\label{sec:conclusion}

In this work, we investigated how searches for GW signals impact the detection of microlensed GW signals, ranging from assessing the impact of using non-adapted searches to demonstrating a new search strategy capable of detecting microlensed signals without requiring a dedicated template bank. We also showed how the search output can be used to differentiate genuinely lensed signals from unlensed triggers. 

We used the \pycbc\ search pipeline with O4 configuration and performed searches on a population of point-mass microlensed signals injected in a 2-detector network and spanning the lens parameter space. We showed that two effects impact the detection of a microlensed signal: (a) the fitting factor of the templates themselves, leading to a loss in signal-to-noise ratio, and (b) the signal-consistency test, which discards signals with a time-frequency evolution that is inconsistent with the template. The two effects lead to different practical impacts on the detection efficiency. While signals with a lower SNR can still be found in the trigger list, though with smaller significance, signals failing the consistency test are fully discarded and never considered. 

We focus on adapting the consistency test to accommodate microlensed signals, which is the main source of loss in detection efficiency. Therefore, we start from the observation that, under the geometric-optics approximation, the signal can be represented by the overlap of two signals with a time delay, a magnification factor, and an overall phase shift. So, one does not need to re-filter the data with a new template bank but rather adapt the consistency test to account for the double-template nature of the signal. Using this adapted statistic, we test two methods to search for microlensed signals: (i) using only the template found in the search, recompute the $\chi^2$ statistic for the lensed hypothesis, and (ii) re-maximize the adapted statistic over a reduced bank. Each of the methods improves the detection efficiency of microlensed signals, but the second method is more effective, providing results similar to the case where the lens statistic is computed with the exact template and lens parameters. 
Using these adapted methods, the relative detection volume is $1.04$ [$1.75$] for (i) and $1.06$ [$1.93$] for (ii) over the full parameter space [for the $10\%$ of injections most affected by lensing].

The method above enables one to uncover triggers and assess their significance compared to noise using time slides. While this ascertains the astrophysical nature of the trigger, it does not assess the lensing nature of the signal. Here, we also propose a method to identify genuinely lensed signals based on the search outputs. Therefore, we built a statistic comparing the squared SNR of the signals recovered under the lensed and unlensed hypotheses, corresponding to twice the difference of the maximum likelihood between the two hypotheses. By then comparing the value found for lensed and unlensed signals, we show that the statistic is able to distinguish between the two populations, with more than 50\% chance of detection for SNRs higher than 15. This method can be used to assess the lensing nature of a trigger, and it can be adapted to other search pipelines. Such an approach enables one to alleviate limitations in doing background calculations to evaluate a trigger's significance where parameter estimation can be prohibitively expensive. 

Additionally, here, we have focused on the case where the two images are contained within one frame. Extending our approach to the case of images separated by larger time delays is straightforward, and it can be done by using the same adapted statistic to search for a second signal in the data. In this case, however, one would also need to include information about the lens population statistic to avoid an important increase in the false alarm rate. Another aspect that would require careful consideration is the generalization of this method to other lens models. For more complex lens models, the geometric optics approximation can lead to more images being formed and wave-optics effects can lead to more complex interference patterns. In this case, the adapted statistic may need to be generalized to more images, in which case it will suffer from increased trials factors, or the geometric optics modeling may be insufficient to capture the features in the data. Should this be the case, then an adapted template bank may be the only way to recover those signals. Such considerations need to be tested more carefully in the future, using a complete search workflow, similar to this work. The same holds for a microlens embedded in a macro-lens potential. The local convergence and shear change the magnifications, and additional images can be formed.

By combining the search part and the lensing identification aspect, this work provides an end-to-end method to search for microlensed signals in GW data and quantify their statistical significance without requiring the use of a dedicated template bank or the use of parameter estimation techniques. Therefore, it is also capable of tackling the problems of searching for microlensed signals in low-latency, and can be deployed on the very large backgrounds needed to have a significant detection of a lensed GW signal.

\begin{acknowledgments}
We thank Anand Sengupta for pointing out an error in our estimate of the search reach. We thank Alvin Ka Yue Li for comments and feedback, Mick Wright for explanations of \gravelamps, and Isaac C. F. Wong for suggesting extra testing to show the efficiency of the search. This work was supported by the Fonds de la Recherche Scientifique (F.R.S.-FNRS) under the funding IISN-4.4501.19. JJ also acknowledges support from the Royal Observatory of Belgium. We have used \numpy~\cite{Harris:2020xlr}, \scipy~\cite{Virtanen:2019joe}, \matplotlib~\cite{Hunter:2007ouj} for analyses and preparing the figures in the manuscript. The authors are grateful for computational resources provided by the LIGO Laboratory and supported by the National Science Foundation Grants PHY-0757058 and PHY-0823459.

\end{acknowledgments}

\appendix

\begin{figure}
    \centering
    \includegraphics[width=\linewidth]{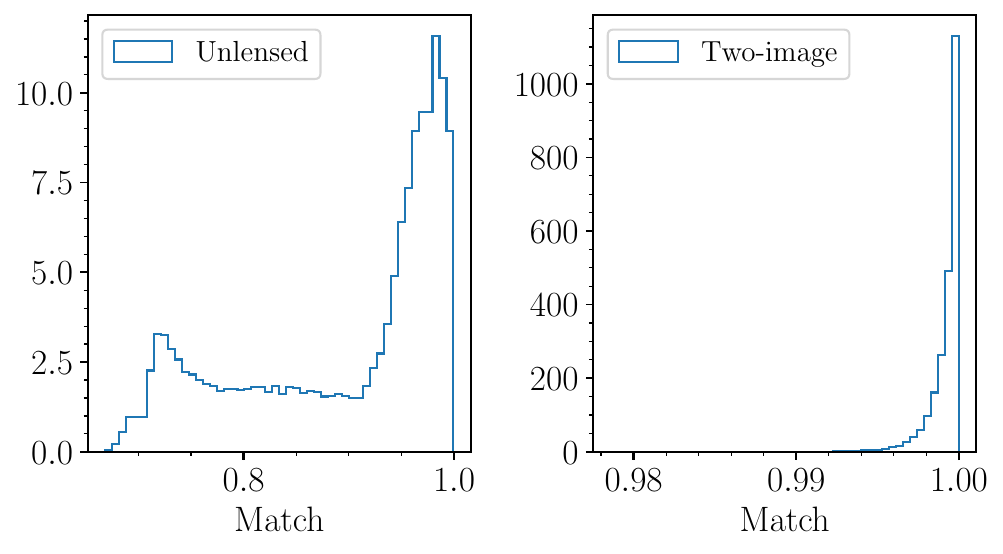}
    \caption{Distribution of the match for 50,000 random binary and lens parameters. Left: match between the lensed waveform and the unlensed waveform of the same binary. Right: match between the lensed waveform and the two-image model, optimized over the time delay and the complex coefficients. Note the different $x$-axis ranges. The unlensed matches spread down to about $0.7$, while $99.9\%$ of the two-image matches are above $0.99$.}
    \label{fig:valid_two_image}
\end{figure}

\section{Validation of the two-image approach}
\label{app:validation}

Here we test how well the two-image approach describes a lensed signal. We do this without noise, so the result shows only the efficiency of the approximation.

We draw 50,000 random points for the binary and lens parameters, using the same distributions as in Table~\ref{tab:ParamSpace}. For each point, we first compute the match between the lensed waveform and its corresponding unlensed waveform, maximizing over time and phase. This quantifies the mismatch introduced by lensing for the same intrinsic binary parameters. We then compute the match with the two-image model, constructed from the unlensed waveform and its time-shifted copy. The complex coefficients of the two images are maximized analytically, allowing arbitrary relative magnification and phase, while the time delay is optimized over a grid.

Fig.~\ref{fig:valid_two_image} shows the distributions of the two match values. The unlensed template often fails to describe the signal. The match spreads from about 0.7 to 1, with a long tail. In contrast, the two-image template recovers the lensed waveform very well. In contrast, the two-image template recovers the lensed waveform very well. The match is above $0.99$ for $99.9\%$ of the points, and above $0.999$ for most of them. So the two-image approach captures almost all of the lensing effect in this parameter space.

\clearpage

\bibliographystyle{apsrev4-2-titles}
\bibliography{reference}

\end{document}